\documentclass[reprint,superscriptaddress,nofootinbib]{revtex4-2}
\usepackage{amsmath,amssymb}
\usepackage{mathptmx}
\usepackage{hyperref}
\usepackage{cleveref}
\usepackage{xurl}
\usepackage{graphicx}
\usepackage{booktabs}
\usepackage{tabularx}
\usepackage[ruled]{algorithm2e}
\usepackage{xcolor}

\newcommand{\rbn}{\ensuremath{(\mathbf{B}\cdot\mathbf{\hat{n}})/|\mathbf{B}|}}
\newcommand{\mrbn}{\ensuremath{\langle |\mathbf{B}\cdot\mathbf{\hat{n}}|/|\mathbf{B}| \rangle}}

\begin{document}

\title{A framework for discrete optimization of stellarator coils}

\author{K.~C.~Hammond}
\email{khammond@pppl.gov}
\affiliation{Princeton Plasma Physics Laboratory, Princeton, NJ, USA}

\begin{abstract}

Designing magnets for three-dimensional plasma confinement is a key task for
advancing the stellarator as a fusion reactor concept. Stellarator magnets
must produce an accurate field while leaving adequate room for other 
components and being reasonably simple to construct and assemble.
In this paper, a framework for coil design and optimization is introduced
that enables the attainment of sparse magnet solutions with arbitrary 
restrictions on where coils may be located.
The solution space is formulated as a ``wireframe'' consisting of
a mesh of interconnected wire segments enclosing the plasma. Two methods
are developed for optimizing the current distribution on a wireframe:
Regularized Constrained Least Squares (RCLS), which uses a 
linear least-squares approach to optimize the currents in each segment,
and Greedy Stellarator Coil Optimization (GSCO), a fully discrete
procedure in which loops of current are added to the mesh one by one to 
achieve the desired magnetic field on the plasma boundary. Examples
are presented of solutions obtainable with each method, 
some of which achieve high field accuracy while obeying spatial
constraints that permit easy assembly.

\end{abstract}

\maketitle

\section{Introduction}

Stellarators aim to confine hot, dense plasmas using a magnetic field generated
entirely by external magnets, without the need to drive current within the
plasma. For this concept to work successfully, the magnetic field must be
carefully shaped in order to ensure that plasma particles and energy are well
confined \cite{imbert-gerard2025a}. 
Designing magnets to produce a field that meets these shaping
requirements typically requires numerical optimization. Optimizing stellarator 
magnets involves three interconnected components: (1) an objective
function that quantifies how well the desired properties have been achieved
by a given design, (2) a framework that maps a set of numerical parameters to a 
magnet design, and (3) an algorithm that adjusts the design parameters in 
order to find the best possible value of the objective function. 

With regard to the second component, magnet optimization problems provide
extensive flexibility because there are infinitely many distributions of 
external sources that can produce a given magnetic field in the vicinity of the
plasma \cite{landreman2017a}. In most stellarator designs these sources are 
electromagnetic coils, although other sources are possible, including permanent 
magnets \cite{helander2020a} and superconducting monoliths \cite{bromberg2011a}.
For an optimization problem to be tractable, the design framework
can only sample a portion of this infinite solution space. Correspondingly,
many such frameworks are possible, each one encompasing different portions
of the space.

One commonly-used framework entails the definition of a toroidal 
\textit{winding surface} that encloses the plasma and carries a surface
current distribution. The surface current can be represented by a scalar 
current potential function, which is most often parametrized with Fourier 
coefficients. These parameters can be optimized with a linear least-squares
approach in codes such as \textsc{Nescoil} \cite{merkel1987a} or
\textsc{Regcoil} \cite{landreman2017a}.
The level curves of the optimized current 
distribution can form the basis for the design of coils. 
While the standard application of this framework holds
the shape of the winding surface fixed, further improvements can be gained
by optimizing the winding surface itself and/or by directly optimizing the 
paths taken by the coils along the surface \cite{drevlak1998a, miner2000a,
paul2018a}. 

In another framework, the field sources are a set of current-carrying space
curves that directly represent the shapes of coils. The space curves may
be parametrized, for example, with Fourier coefficients \cite{strickler2002a,
zhu2018a} or spline knots \cite{brown2015a, yamaguchi2021a, lonigro2022a}. 
The fidelity of the framework to realistic coil design may be improved by
modeling coils with finite dimensions \cite{singh2020a, mcgreivy2021a}.
Space curve parameterizations have been optimized with nonlinear quasi-Newton
codes such as \textsc{Focus} \cite{zhu2018a} or a genetic algorithm as
in \textsc{Gospel} \cite{yamaguchi2021a}.
Compared to the winding surface approach, the space curve approach has the
advantage that the geometry of each coil can be independently parametrized,
and optimizers can more easily explore the three-dimensional volume
around the plasma. 
On the other hand, the space curve approach has less topological flexibility,
as the number of coils must be predefined.

The exploration of permanent magnet stellarator designs has led to the 
development of new frameworks for parametrization and optimization. In some 
cases, the concept of the winding surface is extended to represent a surface of
magnetized material of varying thickness. Such surfaces have been 
optimized using iterated linear least-squares procedures 
\cite{zhu2020a, landreman2021a, xu2021a}.
In other frameworks, the solution space consists of an array of independently 
parametrized magnetic dipoles with arbitrary spatial positions, for which
the dipole moments are then optimized using nonlinear continuous procedures
\cite{zhu2020b, kaptanoglu2022a}, and subsequently discretized to minimize
the number of unique magnet types \cite{hammond2022a, qian2023a, madeira2024a}. 
Magnetic dipole arrays have also been constructed and optimized using 
discrete greedy algorithms \cite{lu2021a, lu2022a, kaptanoglu2023a,
hammond2024a}. 

The representation of permanent magnet solutions as arrays of independent
dipoles illustrated the advantages of a spatially local parametrization,
which made it easier to design magnet distributions around specific obstacles
and optimize for sparsity. More recently, spatially local parametrizations
have been explored for distributions of current density as well. Examples 
include current potential patches defined on 2D winding 
surfaces \cite{elder2024a} and current voxels that fill arbitrary 3D 
volumes \cite{kaptanoglu2023b}.

This paper introduces a new framework for optimizing discrete distributions of 
current with a spatially local parametrization. The framework, hereafter 
referred to as a \textit{wireframe}, consists of an interconnected network
of current-carrying wire segments. The parameters for the distribution are 
simply the currents within each segment; therefore it is straightforward to 
block off segments lying in arbitrary regions of design space to leave room 
for other components.
While examples in this paper all use wireframes that exhibit two-dimensional
topology similar to that of a winding surface, wireframes can in principle be 
designed to fill arbitrary three-dimensional volumes.

This framework enables multiple optimization strategies, including a 
rapid linear least-squares approach and a fully discrete greedy approach
analogous to greedy algorithms previously used with permanent magnets.
As will be demonstrated, the greedy optimizer in particular can be utilized
to obtain sparse solutions with arbitrary spatial constraints, yielding
starting points for convenient coil designs that could not be easily found
using previous methods.

The paper will be organized as follows. Section \ref{sec:wireframe} will
introduce the concept of the wireframe in detail along with the 
governing equations for calculating the field and setting constraints
on its current distribution. Section \ref{sec:rcls} will describe the 
linear Regularized Constrained Least Squares (RCLS) optimization method
along with some example solutions that achieve high field accuracy on a 
relatively coarse wireframe. Section \ref{sec:gsco} will describe the
discrete Greedy Stellarator Coil Optimization (GSCO) algorithm and 
present a number of basic use cases that lead to coil solutions with 
potentially convenient properties.

\section{The wireframe}
\label{sec:wireframe}

The wireframe as defined in this paper is a parametrized solution space for a 
spatial distribution of electric current. Fundamentally, it consists of a set
of interconnected straight segments of filamentary, current-carrying ``wire''. 
Segments connect to one another
at \textit{nodes}, and in turn, the location and orientation of each segment
is defined by the positions of the two nodes that form its endpoints. Each
segment carries a certain current. Thus, the spatial current distribution in a 
wireframe containing $N$ segments can be fully specified by the positions of its
nodes, the pairs of nodes that each segment is connected to, and the current
carried by each segment. 

\begin{figure}
    \begin{center}
    \includegraphics[width=0.48\textwidth]{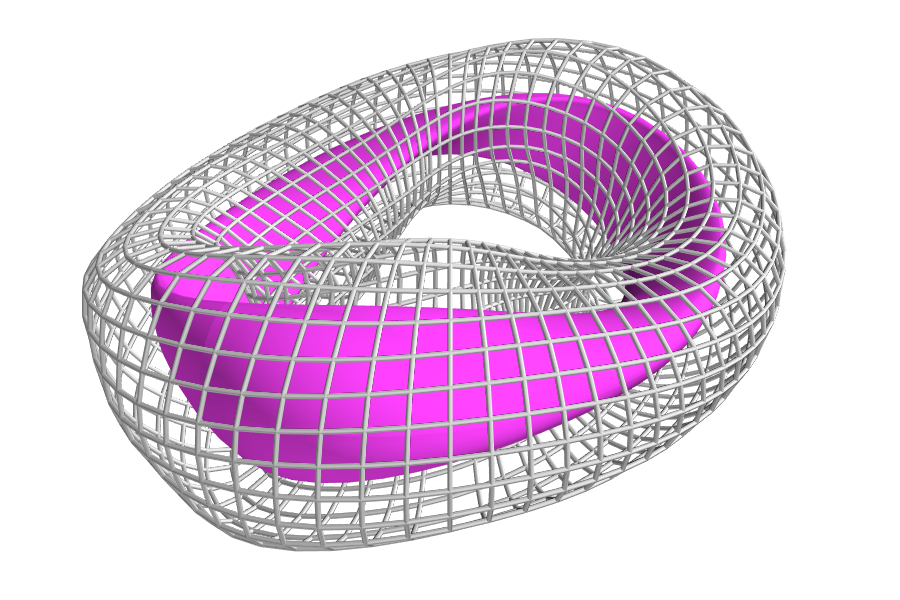}
    \caption{Example of a wireframe with toroidal topology (gray) 
             enclosing a plasma (magenta).}
    \label{fig:wframe_schematic}
    \end{center}
\end{figure}

An example of a wireframe is depicted in Fig.~\ref{fig:wframe_schematic}.
In this case, the segments of the wireframe join to form a toroidal mesh
that encloses a stellarator plasma. The wireframe is constructed by first
defining a toroidal reference surface around the plasma. A 2D grid of points
spaced regularly in the toroidal and poloidal angles is then placed on 
the reference surface; these points are defined as the positions of the nodes. 
The segments are then constructed along straight lines that join adjacent
nodes in the toroidal and poloidal grid dimensions. Segments that connect
adjacent nodes in the toroidal dimension will be referred to as 
\textit{toroidal segments}, whereas segments that connect adjacent nodes
in the poloidal dimension will be referred to as \textit{poloidal segments}.
The wireframes used in this paper will 
all have toroidal topology, although in general wireframes can have other
topologies. Furthermore, the wireframes studied here will exhibit stellarator
symmetry; thus, for a stellarator with $N_{fp}$ toroidal field periods, a
wireframe is fully specified by the segments and currents within one
half-period.

The magnetic field at a point $\mathbf{p}$ due to a wireframe segment between
nodes located at points $\mathbf{p}_1$ and $\mathbf{p}_2$ can be derived from 
the Biot-Savart law for a filamentary current and is given by 
\cite{anderson1976a,hanson2002a}

\begin{align}
\label{eqn:segment_Bfield}
    \mathbf{B}
        &= \frac{\mu_0 I}{4\pi}
           \left(\frac{|\mathbf{r}_1| + |\mathbf{r}_2|}
                    {|\mathbf{r}_1||\mathbf{r}_2|
                        (|\mathbf{r}_1||\mathbf{r}_2| 
                             + \mathbf{r}_1 \cdot \mathbf{r}_2)} \right)
           \mathbf{r}_1 \times \mathbf{r}_2,
\end{align}

\noindent where $\mu_0$ is the vacuum permeability constant, $I$ is the current 
carried by the segment with positive current defined as flowing from 
$\mathbf{p}_1$ to $\mathbf{p}_2$, $\mathbf{r}_1 = \mathbf{p} - \mathbf{p}_1$,
and $\mathbf{r}_2 = \mathbf{p} - \mathbf{p}_2$. A derivation of this expression
is provided in Appendix \ref{apx:segment_field_derivation}.

An important quantity for evaluating the suitability of an externally-generated
magnetic field to confine a stellarator plasma is the normal component of the
field on the boundary of the desired plasma. Following 
Eq.~\ref{eqn:segment_Bfield}, the normal component $B_{ni}$ of the 
magnetic field from the wireframe at a test point $\mathbf{p}_{i}$ on the 
plasma boundary can be expressed as a linear combination of the currents in the 
wireframe's $N$ segments,

\begin{align}
    \label{eqn:bnormal}
    B_{ni} &= \sum_{j=1}^{N} G_{ij}x_{j},
\end{align}

\noindent where $x_{j}$ is the current in the $j^\text{th}$ segment of the 
wireframe and $G_{ij}$ contains the geometric terms relating the current
to the normal field component:

\begin{align}
\label{eqn:G_matrix_element}
    G_{ij} = \frac{\mu_0}{4\pi} 
              \left(\frac{|\mathbf{r}_{1ij}| + |\mathbf{r}_{2ij}|}
                       {|\mathbf{r}_{1ij}||\mathbf{r}_{2ij}|
                           (|\mathbf{r}_{1ij}||\mathbf{r}_{2ij}| 
                             + \mathbf{r}_{1ij} \cdot 
                                   \mathbf{r}_{2ij})} \right)
              \left(\mathbf{r}_{1ij} \times \mathbf{r}_{2ij}\right)
                  \cdot \mathbf{\hat{n}}_{i}
\end{align}

\noindent Here, $\mathbf{\hat{n}}_{i}$ is the unit normal vector at the 
$i^\text{th}$ test point on the plasma boundary; while
$\mathbf{r}_{1ij} = \mathbf{p}_{i} - \mathbf{p}_{1j}$ and
$\mathbf{r}_{2ij} = \mathbf{p}_{i} - \mathbf{p}_{2j}$, 
where $\mathbf{p}_{1j}$ and $\mathbf{p}_{2j}$ are the endpoints
of the $j^\text{th}$ segment.

The methods introduced in this paper seek to optimize the current distribution 
in the wireframe while keeping its geometry fixed. Therefore, the optimization
problem consists of finding an optimal set of currents $\mathbf{x}$ for each
segment in the wireframe. 

While $\mathbf{x}$ contains $N$ elements -- one 
current for each segment -- in general the problem has fewer than $N$ degrees
of freedom due to constraints on the current distribution. For example, 
it is important to impose constraints in order to avoid charge accumulation at 
the nodes. In other words, the current flowing into each node should
equal the current flowing out. For a given node with the label $k$, this can
be expressed as a linear equality constraint:

\begin{equation}
    \label{eqn:constr_continuity}
    \sum_{j \in S_{+k}} x_j - \sum_{j \in S_{-k}} x_j = 0,
\end{equation}

\noindent where $S_{+k}$ is the set of segments for which positive current flows
inward toward node $k$ and $S_{-k}$ is the set of segments for which positive
current flows outward from node $k$. These constraints will be referred to as 
\textit{continuity constraints} because they impose current continuity and
prevent charge accumulation within the wireframe.

\begin{figure}
    \begin{center}
    \includegraphics[width=0.48\textwidth]{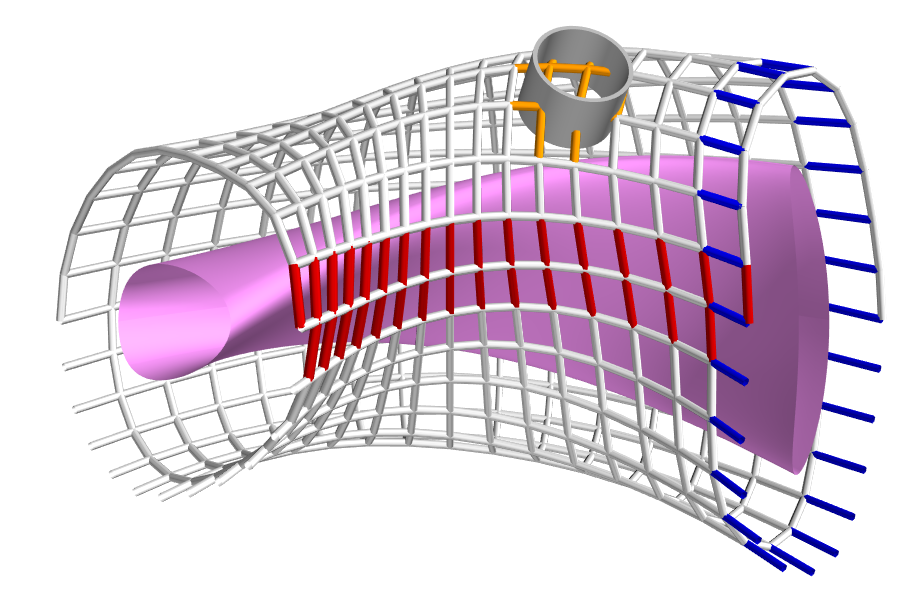}
    \caption{Schematic of one half-period of a wireframe along with the plasma
             boundary in magenta and an example diagnostic port in the form
             of the gray tube. Segments in $S_{pol}$ 
             (Eq.~\ref{eqn:constr_pol_cur}) relevant to the poloidal current 
             constraint are shown in red. Segments in $S_{tor}$ 
             (Eq.~\ref{eqn:constr_tor_cur}) relevant to the toroidal current
             constraint are shown in blue. Segments constrained individually
             to carry zero current (i.e. to avoid collisions with the port)
             are shown in orange.}
    \label{fig:wframe_constraints}
    \end{center}
\end{figure}

Additional constraints may be imposed to ensure that current distributions
exhibit certain desired properties. For example, a \textit{poloidal current
constraint} imposes a value for the net poloidal current $I_{pol}$ flowing 
within the wireframe. This ensures that average toroidal magnetic field $B_t$ 
at a given major radius $R$ within the wireframe has the value 
$\mu_0I_{pol}/(2\pi R)$. Assuming that the continuity constraints of 
Eq.~\ref{eqn:constr_continuity} are in effect, this constraint can be formulated
by requiring that the sum of the currents flowing through all of the poloidal 
segments within a row extending toroidally around the wireframe grid equal 
$I_{pol}$, as shown by the red segments in Fig.~\ref{fig:wframe_constraints}. 
This constitutes one additional linear equality constraint,

\begin{equation}
    \label{eqn:constr_pol_cur}
    \sum_{j \in S_{pol}} x_j = I_{pol},
\end{equation}

\noindent where $S_{pol}$ contains the indices of the poloidal segments.
While it is sufficient in principle to apply the constraint to the sum the 
segment currents in a single toroidal row, two rows must be included in this 
case to uphold the stellarator symmetry of the current distribution.

Similarly, a \textit{toroidal current constraint} may be imposed to require 
that the wireframe carry a certain net current $I_{tor}$ in the toroidal 
dimension. Setting $I_{tor}$ to zero prevents current from 
flowing helically around the torus, ensuring that the wireframe has no net
dipole moment. On the other hand, if $I_{tor}$ and $I_{pol}$ are both set to
specific values, this  will enforce a certain helicity in the current 
distribution. Assuming the continuity constraints 
(Eq.~\ref{eqn:constr_continuity}) apply, $I_{tor}$ can be constrained by 
ensuring that the sum of the currents through the toroidal segments in one
poloidal column equals $I_{tor}$, as shown by the blue segments in 
Fig.~\ref{fig:wframe_constraints}. This constitutes another linear equality
constraint,

\begin{equation}
    \label{eqn:constr_tor_cur}
    \sum_{j \in S_{tor}} x_j = I_{tor},
\end{equation}

\noindent where $S_{tor}$ contains the indices of the toroidal segments in one 
poloidal column. 

Finally, it may be desired to prevent any current from flowing within a given
segment. This would be the case, for example, if the segment lies in a spatial
region that is reserved for a component such as a port for plasma diagnostics 
or heating systems. Such a \textit{segment constraint} for a given segment
with index $j$ may be expressed simply as

\begin{equation}
    \label{eqn:constr_segment}
    x_j = 0
\end{equation}

\noindent An example set of constrained segments is shown in orange in 
Fig.~\ref{fig:wframe_constraints} to avoid the presence of currents in 
the vicinity of a port.

Since the constraints described above are all linear equality constraints,
any combination of these constraints applicable to a given wireframe may be
expressed comprehensively as a system of linear equations,

\begin{equation}
    \label{eqn:constr_all}
    \mathbf{C}\mathbf{x} = \mathbf{d},
\end{equation}

\noindent where $\mathbf{C}$ is a matrix containing the coefficients of $x_j$ 
from the left-hand sides of Eq.~\ref{eqn:constr_continuity},
\ref{eqn:constr_pol_cur},\ref{eqn:constr_tor_cur},\ref{eqn:constr_segment} and 
$\mathbf{d}$ contains the terms on the right-hand sides of these equations.
Formulating the constraints in this way enables the use of constrained linear
optimization procedures, such as the one described in Sec.~\ref{sec:rcls}.

\section{Regularized Constrained Least Squares (RCLS)}
\label{sec:rcls}

The current distribution in the wireframe can be optimized with many different
techniques, each yielding solutions with distinct characteristics and features.
This paper will describe two such optimization methods, each of which works
to minimize an objective function $f$ through adjustment of the currents 
$\mathbf{x}$ in the segments of the wireframe.
The first method, described in this section, will be called \textit{Regularized
Constrained Least Squares} (RCLS). It utilizes a linear least-squares 
technique that has the advantages of being rapid and non-iterative.

\subsection{Description of the method}
\label{sec:rcls_desc}

For a linear least-squares solver to be applicable, the objective function
$f$ must be quadratic in the parameters to be optimized, in this case 
$\mathbf{x}$. For the RCLS procedure described in this section the objective
function $f_{RCLS}$ is defined as a sum of two sub-objectives:

\begin{equation}
    \label{eqn:f_rcls}
    f_{RCLS} = f_B + f_R
\end{equation}

The first sub-objective, $f_B$, is a metric of magnetic field error commonly
used in stellarator magnet optimization. It is related to the square of
the discrepancy from the desired values of the normal component of the magnetic 
field on the target plasma boundary produced by the wireframe:

\begin{equation}
    \label{eqn:f_B}
    f_B = \frac{1}{2}\left(\mathbf{A}\mathbf{x}-\mathbf{b}\right)^2 
\end{equation}

\noindent Hereafter, the square of a vector quantity constitutes the dot product
of the vector with itself. $\mathbf{A}$ contains the geometric matrix elements 
from $G$ as defined in Eq.~\ref{eqn:G_matrix_element} weighted by the square 
root of the area $a_i$ of the portion of the target plasma boundary attributed 
to the respective test point:

\begin{equation}
    \label{eqn:A_matrix_element}
    A_{ij} = G_{ij} \sqrt{a_i}
\end{equation}

\noindent $\mathbf{b}$ contains the desired values of the normal magnetic field 
at test points on the target plasma boundary, also weighted by $\sqrt{a_i}$:

\begin{equation}
    \label{eqn:b}
    b_i = \left(\mathbf{B}_{\text{targ},i} \cdot \mathbf{\hat{n}}_i \right)
                 \sqrt{a_i}
\end{equation}

\noindent The target values of the magnetic field, $\mathbf{B}_{\text{targ},i}$
at each test point on the plasma boundary are the values that the field from
the wireframe must attain in order to confine the target plamsa, accounting
for fixed contributions from other field sources including external coils and
currents within the plasma.

As formulated in Eq.~\ref{eqn:f_B}, $f_B$ is approximately equal to half of the 
square integral of the field discrepancy over the plasma boundary:

\begin{equation}
    \label{eqn:f_B_approx}
    f_B \approx \frac{1}{2} \iint 
                  \left[ \left( \mathbf{B}_\text{wf} 
                                - \mathbf{B}_\text{targ} \right)
                  \cdot \mathbf{\hat{n}} \right]^2 dA
\end{equation}

\noindent Here, $\mathbf{B}_\text{wf}$ refers specifically to the magnetic 
field produced by the wireframe current distribution.

The second sub-objective, $f_R$, constitutes Tikhonov regularization 
\cite{tikhonov1963a,tikhonov1995a}. It penalizes segment currents that are 
especially large:

\begin{align}
    \label{eqn:f_R}
    f_R &= \frac{1}{2}\left(\mathbf{W}\mathbf{x}\right)^2
\end{align}

\noindent $\mathbf{W}$ is a (typically diagonal) matrix 
of weighting terms. If $\mathbf{W}$ is indeed diagonal and all diagonal elements
are the same, then $f_R$ reduces to the form 
$(1/2)\lambda\mathbf{x}^2$, where $\lambda$ is the square of each diagonal 
element. Assigning different values to the diagonal elements of $\mathbf{W}$,
on the other hand, allows the currents in different segments to be weighted
differently for the regularization.

The RCLS algorithm finds the optimal set of segment currents $\mathbf{x}^*$ 
that minimizes the total objective $f_{RCLS}$ while upholding a set of linear 
equality constraints:

\begin{align}
    \label{eqn:rcls_problem}
    \mathbf{x}^* =& \min_\mathbf{x} \left( f_B + f_R \right) \\
    \label{eqn:rcls_problem_constr}
    &~~ s.t.~~ \mathbf{C}\mathbf{x} = \mathbf{d}
\end{align}

\noindent  $\mathbf{C}$ and $\mathbf{d}$ are the left-hand side coefficients
and right-hand side terms of the constraint equations as defined in 
Eq.~\ref{eqn:constr_all}. 

The solution $\mathbf{x}^*$ to this problem may be found by solving two systems
of linear equations, one with $p$ equations and the other with $n-p$ equations,
where $p$ is the number of constraint equations and $n$ is the number of 
wireframe segments. The procedure is described in more detail in 
Appendix \ref{apx:rcls_derivation}. 

Mathematically, this approach has much in common with the \textsc{Regcoil} code
for designing stellarator coils \cite{landreman2017a}. Both approaches use
linear least squares to minimize the square integral of the normal field
on a target plasma boundary subject to Tikhonov regularization. The difference 
lies in how the (coil) current distribution is formulated and parametrized.
In the case of \textsc{Regcoil}, the parameters are Fourier coefficients of the
current potential of a two-dimensional surface current distribution that 
encloses the plasma. With RCLS, by contrast, the parameters are the wireframe
segment currents and are therefore spatially localized. In addition, in 
\textsc{Regcoil}, current continuity is inherently upheld by the current
potential formulation; whereas with RCLS, a set of linear equality constraints
is required to uphold current continuity.

\begin{figure}
    \begin{center}
    \includegraphics[width=0.5\textwidth]{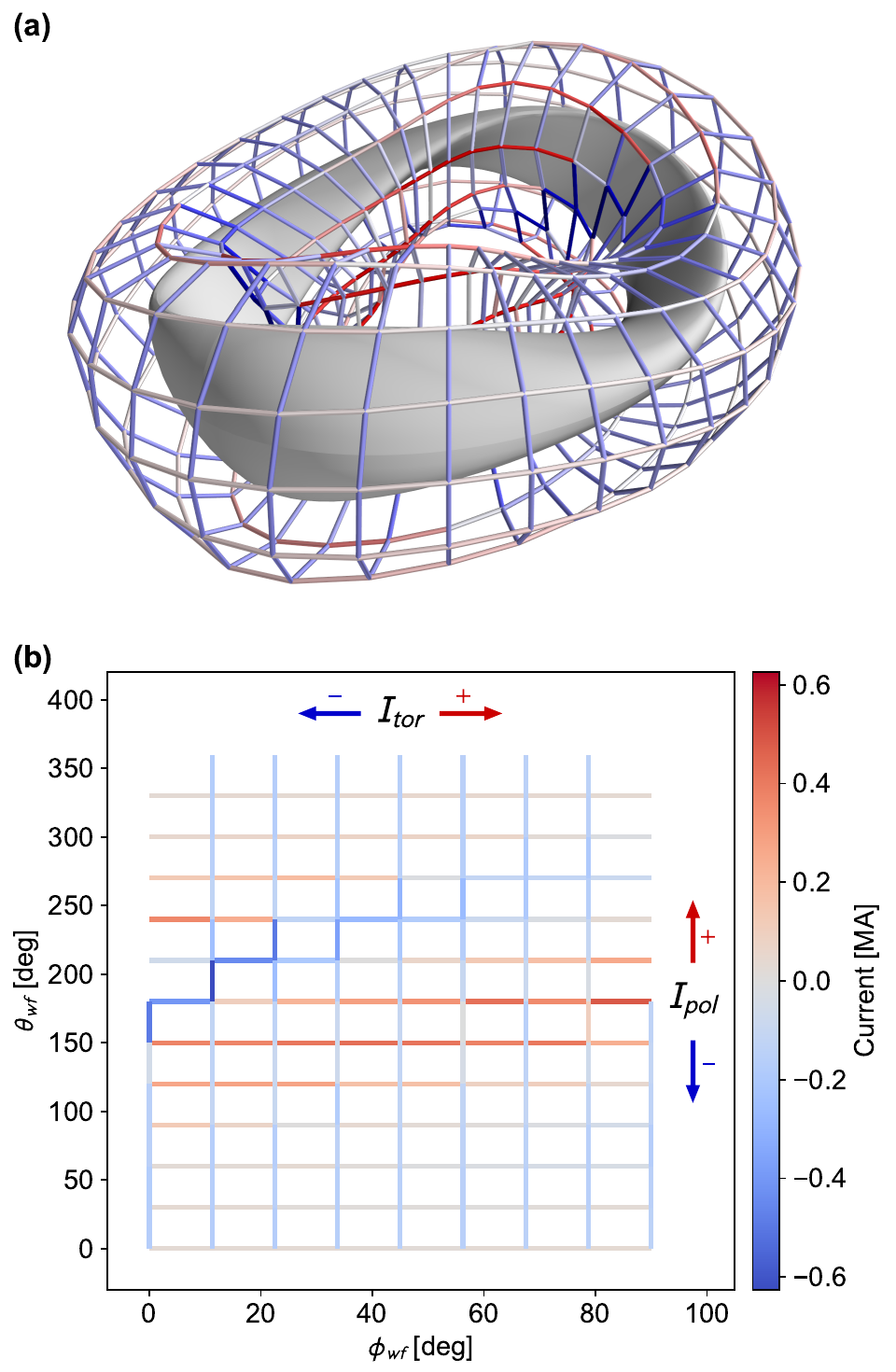}
    \caption{(a) Rendering of a wireframe with currents optimized to produce
                 a field to confine the Precise QA plasma equilibrium described
                 in Ref.~\cite{landreman2022a}, along with the the target 
                 plasma boundary. Segments are color-coded according to the 
                 current they carry.
             (b) Schematic of the segments of one half-period of the wireframe.
                 Arrows on the top of the plot near ``$I_{tor}$'' indicate
                 the current flow direction associated with the colors of the
                 horizontal (toroidal) segments; arrows on the right of the 
                 plot near ``$I_{pol}$'' indicate the flow direction associated
                 with the colors of the vertical (poloidal) segments.}
    \label{fig:wf_rcls_noports}
    \end{center}
\end{figure}

\subsection{Basic example}
\label{sec:rcls_no_ports}

An example of a wireframe with currents optimized by the RCLS procedure
is shown in Fig.~\ref{fig:wf_rcls_noports}. This wireframe encloses the 
stellator equilibrium with near-precise quasiaxisymmetry described in
Ref.~\cite{landreman2022a} (hereafter ``Precise QA''). 
The equilibrium has two field periods, a major radius of 1~m, and a magnetic 
field on axis of 1~T. It is a vacuum equilibrium; hence, the pressure is 
everywhere zero and there are no internal plasma currents.
The wireframe was designed such that the nodes are each approximately 0.3~m
from the target plasma boundary. Segments and nodes lying in
poloidal cross-sections coincide with planes of constant toroidal 
angle. The segments form a toroidal mesh with $N_{tor}$ = 8 nodes per 
half-period in the toroidal dimension (i.e. 32 nodes per toroidal revolution) 
and $N_{pol}$ = 12 nodes in one poloidal revolution. The resolution of this
wireframe will be denoted hereafter as $8 \times 12$ or, more generally,
$N_{tor} \times N_{pol}$. In general, a half-period contains $N_{tor}N_{pol}$ 
toroidal segments and the same number of poloidal segments, leading
to a total of $2 N_{tor} N_{pol}$ segments per half-period.

For this wireframe, the RCLS procedure optimized the currents in the
192 unique segments constituting a single half-period, which by symmetry
specify the currents in all 768 segments within the torus.
The segments' currents were all subject to continuity 
constraints (Eq.~\ref{eqn:constr_continuity}), and a poloidal current
constraint (Eq.~\ref{eqn:constr_pol_cur}) mandated that the net poloidal
current flowing in the wireframe equal 5~MA to produce an average toroidal
field of 1~T at major radius $r=1$~m. Altogether, there were 95 constraint
equations, leaving 97 degrees of freedom for the optimization.
The solution was obtained using a regularization matrix of 
$\mathbf{W}=(10^{-10}$ Tm$/$A$) \mathbf{I}$, where 
$\mathbf{I}$ is the $n \times n$ identity matrix with $n=192$ equal to the
number of segments. The solution took approximately 100 ms to compute on
a laptop computer. All of the equality constraints were very well satisfied,
with the largest component of the residual vector $\mathbf{Cx}-\mathbf{d}$
having an absolute value of less than $1 \times 10^{-9}$~A (for reference,
the smallest current carried in any of the wireframe segments had an absolute
value of $1.7 \times 10^{3}$~A).

Fig.~\ref{fig:wf_rcls_noports}a shows a three-dimensional rendering of the
wireframe and the target plasma boundary. Fig.~\ref{fig:wf_rcls_noports}b
contains a two-dimensional schematic of the segments and currents. The latter
figure depicts the segments from one half-period of the wireframe (one-fourth
of the torus in this case) flattened out with the toroidal dimension on the
horizontal axis and the poloidal dimension on the vertical axis. 
The coordinates correspond to the toroidal $\phi_{wf}$ and poloidal
$\theta_{wf}$ angles at which the nodes appear on the toroidal reference
surface used to construct the wireframe (the subscript ``wf'' is meant to
distinguish the wireframe reference surface angles from the angles 
parametrizing the target plasma boundary).
Toroidal segments, which carry current in (approximately) the toroidal 
dimension, are oriented horizontally; whereas poloidal segments, 
which carry current in (approximately) the poloidal dimension, are oriented
vertically. The plot is periodic in the poloidal (vertical) dimension; hence,
the tops of the poloidal segments appearing at $\theta_{wf}=360^\circ$ are 
connected to the segments appearing at $\theta_{wf}=0^\circ$ on the outboard 
side. The poloidal segments on the left-hand and right-hand sides of the plot 
(toroidal angles $\phi_{wf}=0^\circ$ and $90^\circ$) lie in 
symmetry planes at the interfaces between adjacent half-periods; as such, half 
of the segments in those planes are not shown because they essentially 
``belong'' to the adjacent half-periods in accordance with stellarator symmetry 
\cite{dewar1998a}. 

As indicated by the arrows in the margins of Fig.~\ref{fig:wf_rcls_noports}b, 
the segments are color-coded according to the magnitude and direction of the
current they carry. For the toroidal segments, red indicates current flow
in the positive toroidal direction (right), whereas blue indicates current flow
in the negative toroidal direction (left). Similarly, for the poloidal segments,
red indicates current flow in the positive poloidal direction (up), whereas
blue indicates current flow in the negative poloidal direction (down). 

As can be seen in Fig.~\ref{fig:wf_rcls_noports},
every segment carries a unique current. This is
a consequence of the linear least-squares optimization approach, in which 
the free parameters -- i.e. the segment currents -- are optimized in a 
continuous space. Nevertheless, current continuity is enforced at each 
node through the incorporation of the linear equality constraints in 
Eq.~\ref{eqn:constr_continuity}; hence, there is no charge accumulation at the 
nodes. Also, while no upper bounds are explicitly enforced on the values of the 
segment currents, the incorporation of Tikhonov regularization effectively 
restricts how large these currents can be.

\begin{figure}
    \begin{center}
    \includegraphics[width=0.5\textwidth]{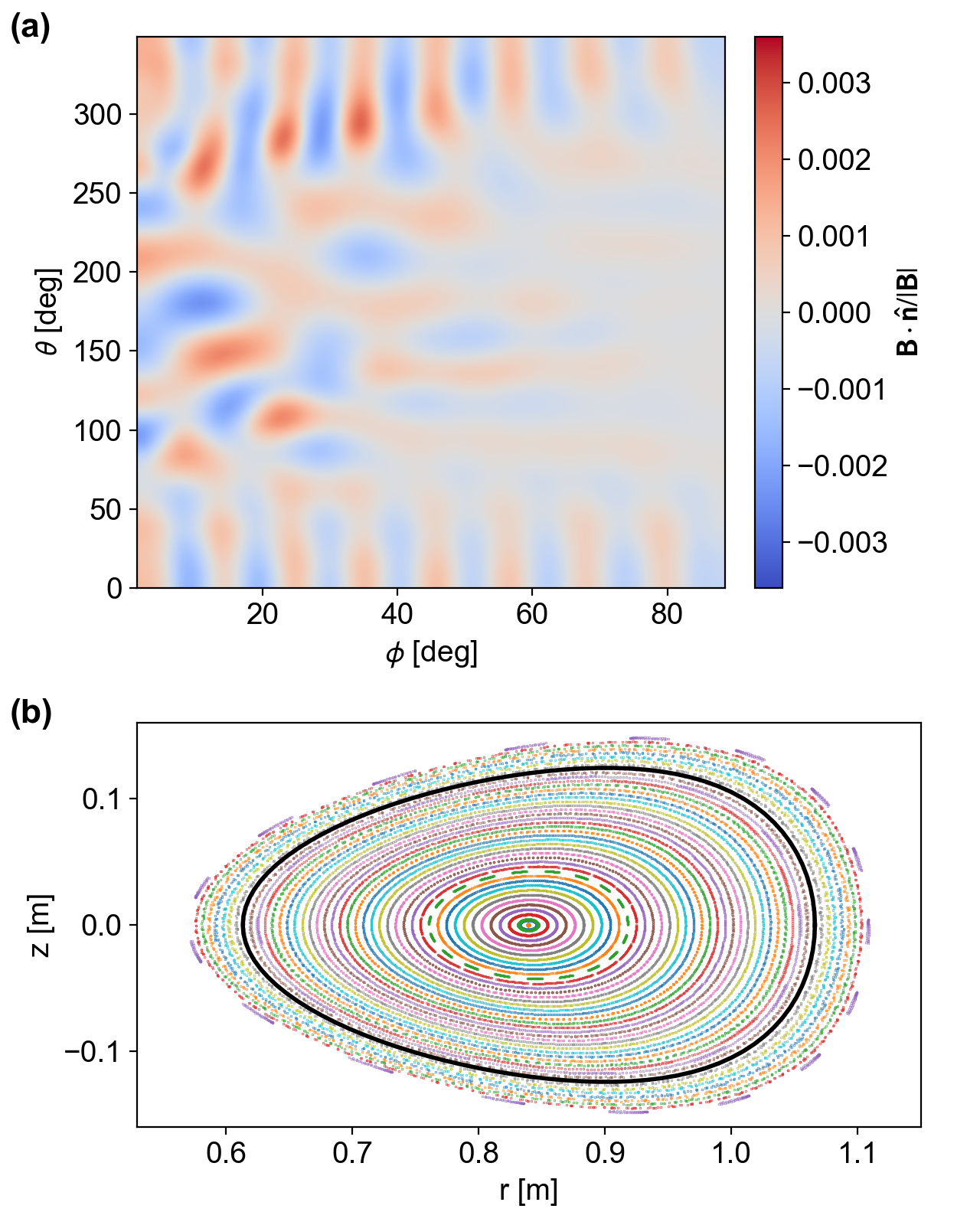}
    \caption{(a) Normal component of the magnetic field on the plasma boundary
                 produced by the wireframe shown in 
                 Fig.~\ref{fig:wf_rcls_noports}, relative to the magnitude
                 of the magnetic field.
             (b) Poincar\'e cross-section of the magnetic field lines produced
                 by the wireframe solution (dots) along with a cross-section
                 of the target plasma boundary (black curve).}
    \label{fig:wf_rcls_noports_field}                
    \end{center}
\end{figure}

Magnetic field accuracy can be quantified by $\rbn$, the relative normal 
component of the total magnetic field $\mathbf{B}$ (including contributions 
from the wireframe as well as plasma currents or external magnets if present)
on the target plasma 
boundary. A perfectly accurate field is tangential to the target plasma 
boundary, and hence $\rbn$ is ideally 0.
The magnetic field produced by this wireframe solution is highly accurate,
as shown in Fig.~\ref{fig:wf_rcls_noports_field}. 
Fig.~\ref{fig:wf_rcls_noports_field}a indicates that \rbn rarely exceeds 0.3\%.
The surface average of the absolute 
value of this normalized quantity over the target plasma boundary, \mrbn, is 
$6.31 \times 10^{-4}$. Furthermore, the flux surfaces produced by the 
solution, whose geometry is indicated by the Poincar\'e cross-sections in 
Fig.~\ref{fig:wf_rcls_noports_field}b, exhibit excellent agreement with the 
shape of the target plasma boundary. 

\subsection{Restricting space for other components}
\label{sec:rcls_ports}

The wireframe formulation makes it straightforward to restrict the solution
to have zero current in arbitrary regions, e.g. to leave space for ports and
other components. As indicated in Fig.~\ref{fig:wframe_constraints}, segments
that overlap restricted areas can simply be constrained to have zero current
by applying segment constraint equations (Eq.~\ref{eqn:constr_segment}).

\begin{figure}
    \begin{center}
    \includegraphics[width=0.5\textwidth]{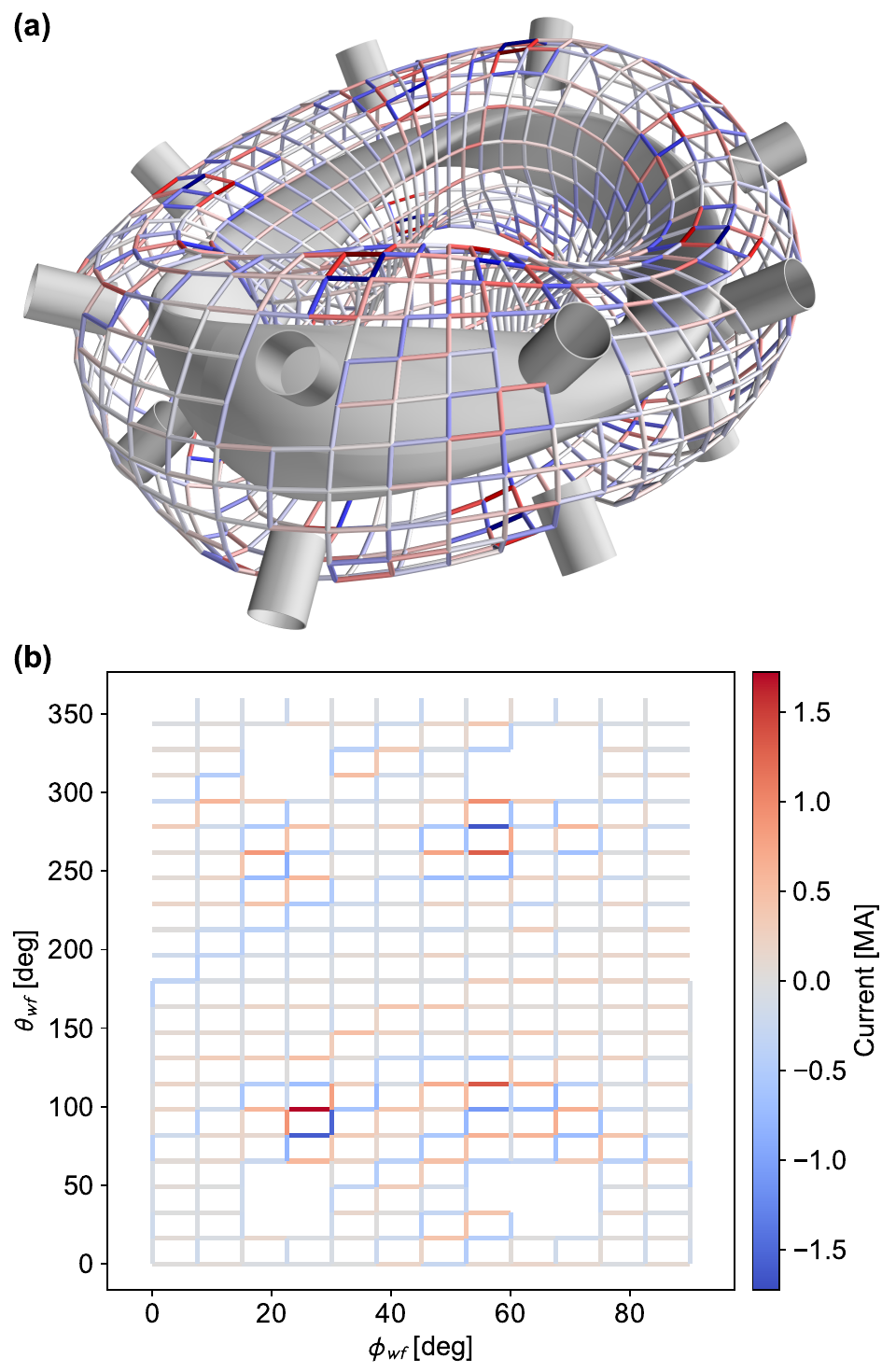}
    \caption{(a) Rendering of a wireframe with a current distribution
                 optimized with the RCLS technique. Segments constrained
                 to have zero current due to overlap with the cylindrical
                 ports are not shown in the rendering.
             (b) Schematic of the segments of one half-period, analogous to 
                 that of  Fig.~\ref{fig:wf_rcls_noports}b, not including 
                 segments constrained to carry no current.}
    \label{fig:wf_rcls_ports}
    \end{center}
\end{figure}

A wireframe solution for the Precise QA plasma with certain areas blocked
off for ports is shown in Fig.~\ref{fig:wf_rcls_ports}. In this example,
each half-period has four cylindrical ports placed on the outboard side. 
To match the accuracy
of the solution without ports as shown in Fig.~\ref{fig:wf_rcls_noports}, it 
was necessary to increase the wireframe grid resolution to $12 \times 22$, 
presumably to compensate for the spatial gaps in the current distribution 
introduced by the segment constraints. Accounting for the constraints placed
on continuity, poloidal current, and the segments that collide with ports,
the wireframe has 241 degrees of freedom to be optimized. For this solution, 
the regularization matrix was $\mathbf{W}=(10^{-10}$ Tm$/$A$) \mathbf{I}$, 
where $\mathbf{I}$ is the $n \times n$ identity matrix with $n=528$ equal to the
number of segments. 

Fig.~\ref{fig:wf_rcls_ports} shows the current distribution in the wireframe
determined for this solution. Segments carrying zero current as a result of
being constrained are omitted. While the overall field accuracy metric
$\mrbn$ is $6.37 \times 10^{-4}$, nearly the same as that of the no-port 
solution, the solutions exhibit some differences.  First, the maximum 
segment current in this solution is roughly twice that of the no-port solution, 
with the largest currents appearing in the vicinity of the holes in the 
distribution at the locations of the ports. Another contrast with the no-port 
solution is in the net toroidal current, which was not constrained for either 
solution. For this solution, the net toroidal current was 0.65 MA, whereas it 
was 0.90 MA in the no-port case. 

It is notable that wireframes with RCLS solutions are able to achieve high
magnetic field accuracy with a relatively coarse grid with few degrees
of freedom, particularly in the no-port case. However, it is not clear how
practical these solutions would be to implement. While each individual straight
segment of the wireframe is conceptually simple, driving unique currents through
the many segments, each of which has junctions with other segments carrying 
different currents, would be challenging.
Nevertheless, RCLS can be useful as a means of rapidly constructing a highly
accurate vacuum magnetic field for a given plasma equilibrium. In addition,
RCLS may be used to optimize the currents in pre-defined paths within the
wireframe, separated from one another with segment constraints that prevent the
existence of multi-current junctions.

\section{Greedy Stellarator Coil Optimization (GSCO)}
\label{sec:gsco}

The RCLS method is just one of many optimization approaches that can be 
employed within the wireframe solution space. This section introduces a new, 
fully discrete optimization technique, hereafter referred to as 
\textit{Greedy Stellarator Coil Optimization} (GSCO). GSCO is an iterative 
method that constructs paths of current within the wireframe that can form the 
basis for the design of coils.

\subsection{Description of the method}
\label{sec:gsco_method}

The GSCO procedure works by adding small loops of current to the wireframe, 
one by one. For the present work, these current loops encircle single
\textit{cells} of the wireframe grid. Each cell consists of four segments that
surround a quadrilateral gap; in other words, the segments forming a cell
do not enclose any other segments. 

\begin{figure}
    \begin{center}
    \includegraphics[width=0.5\textwidth]{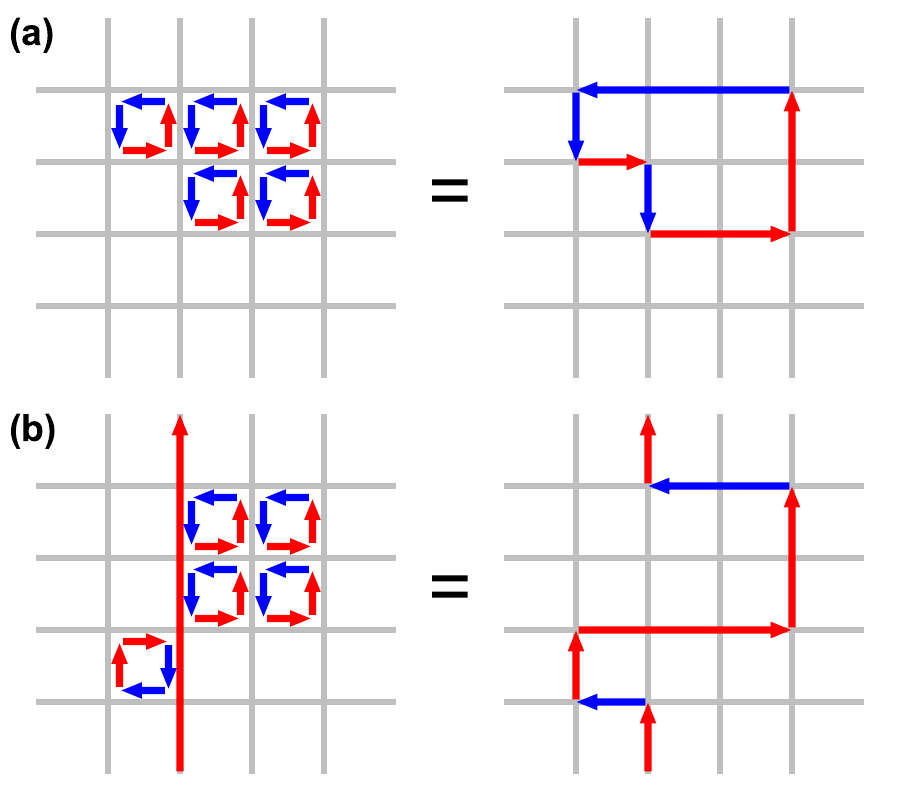}
    \caption{Schematic illustrations of how adding loops of current can 
             effectively add or reshape coils in a wireframe grid. 
             (a) Adding loops of current with the same polarity to 
                 a set of contiguous cells creates a saddle coil.
             (b) Adding loops of current next to a pre-existing straight 
                 section of a coil can add curvature to the coil.}
    \label{fig:loop_schematic}
    \end{center}
\end{figure}

Fig.~\ref{fig:loop_schematic} illustrates schematically how coils may be 
constructed on a wireframe by adding loops of current. 
In Fig.~\ref{fig:loop_schematic}a, five loops with the same polarity are added 
to five contiguous cells of the wireframe. Note that the net current in segments
between adjacent loops is zero, as the contributions from the two loops are
equal and opposite. Accounting for these cancellations, the net effect of the
five loops is to create a saddle coil enclosing the cells where the loops
were added.

Fig.~\ref{fig:loop_schematic}b shows how a pre-existing section of a coil can
be reshaped by adding current loops to cells adjacent to the coil. In this
case, the pre-existing coil is represented by the straight, vertical current
flow. If loops of current are added adjacent to this coil, and the magnitude
and polarity of these loops is such that the loops cancel out the current in
the initially active coil segments, the net effect is to divert the initial 
flow along a new path, as seen in the right-hand plot.

The process of adding loops of current to the wireframe thus constitutes a 
topologically flexible method of designing coils. New coils can be created
by adding loops of current to an empty section of the wireframe, as in 
Fig.~\ref{fig:loop_schematic}a. Existing coils can be reshaped by adding
loops of current to adjacent cells, as in Fig.~\ref{fig:loop_schematic}b.
Furthermore, existing coils can be removed altogether. For example, the coil
created by the five counter-clockwise current loops in 
Fig.~\ref{fig:loop_schematic}a can be removed by adding five clockwise current
loops to the same cells, effectively cancelling out the contributions of 
the original counter-clockwise loops.

The GSCO algorithm optimizes the current distribution in a wireframe by 
performing a series of iterations. In each iteration, a single loop of current 
is added to the wireframe, with the cell (location) and polarity chosen to 
bring about the greatest possible reduction in an objective function 
$f_\text{GSCO}$. The classification of the algorithm as ``greedy'' arises 
from its general approach of constructing a solution to a highly complex
problem in multiple small steps, in each step making an optimal choice 
according to a relatively simple heuristic \cite{aho1983a}.\footnote{
GSCO may also be classified as a repeated \textit{local search}, as the current 
distribution $\mathbf{x}$ is a feasible solution (i.e. all constraints are 
obeyed) at every iteration and thus each iteration entails making
a small change to an already-feasible solution.} Similar 
approaches have been used successfully in many applications including
signal processing \cite{pati1993a}, DNA sequencing \cite{zhang2000a}, and 
operations research \cite{ruiz2007a}.

The GSCO procedure draws particular inspiration from the greedy algorithms 
developed in 
Refs.~\cite{lu2021a,lu2022a,kaptanoglu2023a,hammond2024a} for designing arrays
of permanent magnets to shape the confining fields for stellarator plasmas.
The greedy permanent magnet optimizers (GPMO) worked by adding permanent 
magnets one by one to achieve the desired field shaping. A small loop of 
current in a wireframe functions similarly to a permanent magnet, in the sense
that both constitute sources of approximately dipolar magnetic fields. 
A wireframe current loop can also be thought of as a discrete analogue to the
recently introduced concept of a current potential patch \cite{elder2024a}.
Arrays of small dipole sources can in principle create highly accurate 3D
shaping for stellarator magnetic fields, provided the dipoles are placed in 
the correct locations with the correct orientations and strengths
\cite{helander2020a,zhu2020b}; this insight forms the basis of the heuristic
that guides the iterations in both GPMO and GSCO.

During the greedy iterations, the optimizer faces some limitations on where
it may place loops of current. These limitations are defined by a set of
eligibility rules that dictate whether any given cell in the wireframe
may receive a loop of current. Some possible eligibility rules are summarized
in Table \ref{tab:eligibility}. One essential eligibility rule is that the
wireframe current distribution must adhere to the segment constraints
(Eq.~\ref{eqn:constr_segment}). Clearly, if adding a 
current loop to a given cell would result in a nonzero current appearing in a 
constrained segment, that cell should not be eligible. Other eligibility 
rules are optional and can be used to make the solution simpler. For
example, the user can require that solutions exhibit no crossing current 
paths; i.e. at each node in the wireframe, there can be at most two 
current-carrying segments. With this requirement in place, any cell for which
an added loop of current would result crossed or forked current paths at a node
(i.e. in which more than two segments at a node carry current) is not eligible. 
Note that a cell may be eligible to receive a loop of current with one 
polarity but not the other, depending on the currents initially flowing in
the cell's segments. 

\begin{figure}
    \includegraphics[width=0.5\textwidth]{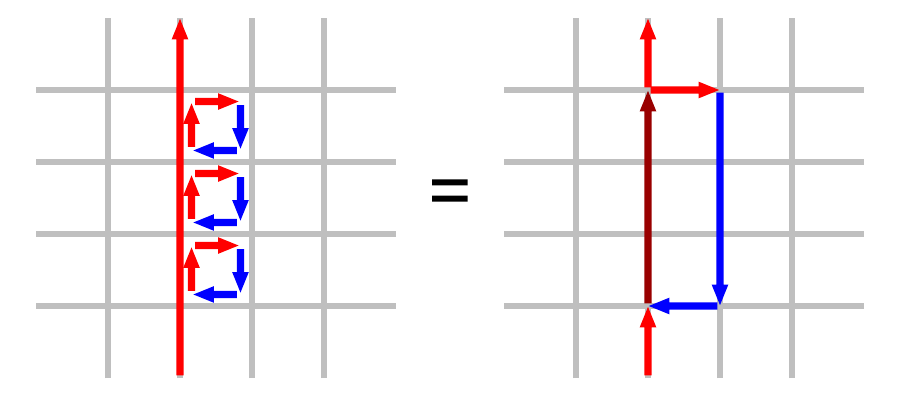}
    \caption{Illustration showing how the addition of loops of current next
             to an existing straight section of wire may create a forked
             current path. In this case, the current loops add to a previously
             existing current flow rather than canceling it out as in 
             Fig.~\ref{fig:loop_schematic}b.}
    \label{fig:forked_paths}
\end{figure}

An example of a loop creating a forked current path
is shown in Fig.~\ref{fig:forked_paths}. In this example, three loops of
current are added next to a straight flow of current. The loops of current
are polarized such that they add to, rather than cancel, the current in the 
segments originally carrying the straight flow. Hence, the net result is
a current distribution in which two segments carrying one unit of current
feed into a section carrying two units of current at the bottom of the figure,
which in turn splits off into two segments carrying one unit of current at the 
top. Such distributions can be undesirable if one is seeking a solution 
with clearly defined, distinct coils whose windings don't merge together or
split apart.

\begin{table}
    \begin{tabularx}{0.5\textwidth}{l l X}
    \toprule
    Rule &~~& Description \\
    \toprule
    wireframe constraints
                        &&  Solution must satisfy all constraint equations
                            $\mathbf{Cx}=\mathbf{d}$ \\
    \midrule
    no crossings        &&  At each node, at most two segments may carry
                            current \\
    \midrule
    no new coils        &&  Loops may not be added to a cell around which all
                            segments presently carry no current \\
    \midrule
    max current ($I_\text{max}$)
                        &&  The absolute value of the current in any segment 
                            may not exceed a given $I_\text{max}$ \\
    \midrule
    max loops per cell ($N_\text{max}$)
                        &&  The net number of positive or negative loops of
                            current added to a given cell may not exceed a
                            defined maximum $N_\text{max}$ \\
    \bottomrule
    \end{tabularx}
    \caption{Eligibility rules that may be applied to determine whether a loop 
             of current may be added to a given cell during a GSCO iteration.}
    \label{tab:eligibility}
\end{table}

The iterations in GSCO proceed until one of two stopping conditions is 
reached. 
The first possible stopping condition is that there are no more cells in the
wireframe that are eligible to have a current loop added. 
The second possible stopping condition is that the optimization has reached a
minimal value of the objective function $f_\text{GSCO}$. In practice, one can
determine that this minimal value has been reached if the optimal current 
loop determined in a given iteration cancels out the current loop added in the
previous iteration. This implies that any subsequent iterations would simply
oscillate between two solutions by adding and removing the same current loop.
Reaching this stopping condition does not necessarily imply that the absolute
minimum of the objective function has been found; rather, the optimizer has
become ``trapped'' in a local minimum of the optimization space.

The objective function $f_{GSCO}$ to be minimized by the GSCO procedure takes
weighted contributions from two sub-objectives:

\begin{equation}
    f_\text{GSCO} = f_B + \lambda_S f_S
\end{equation}

\noindent Here, $f_B$ is the square integral of the normal magnetic field at
the plasma boundary as defined in Eq.~\ref{eqn:f_B}. $\lambda_S$ is a weighting
factor. The second sub-objective $f_S$ is defined as follows:

\begin{equation}
    f_S = \frac{1}{2} N_\text{active},
\end{equation}

\noindent where $N_\text{active}$ is simply the number of segments in the 
wireframe that are active (i.e. they carry nonzero current). Solutions
with lower values of $f_S$ are more sparse than solutions with higher values;
thus the objective function incentivizes sparsity in the solution. 
Sparse solutions are desirable because they are typically simpler to implement
and leave more design space available for other components. Optimizations
with greater values of $\lambda_S$ will prioritize sparsity over field 
accuracy, whereas optimizations with lower values of $\lambda_S$ will prioritize
accuracy over sparsity.

\begin{figure}
    \begin{center}
    \includegraphics[width=0.5\textwidth]{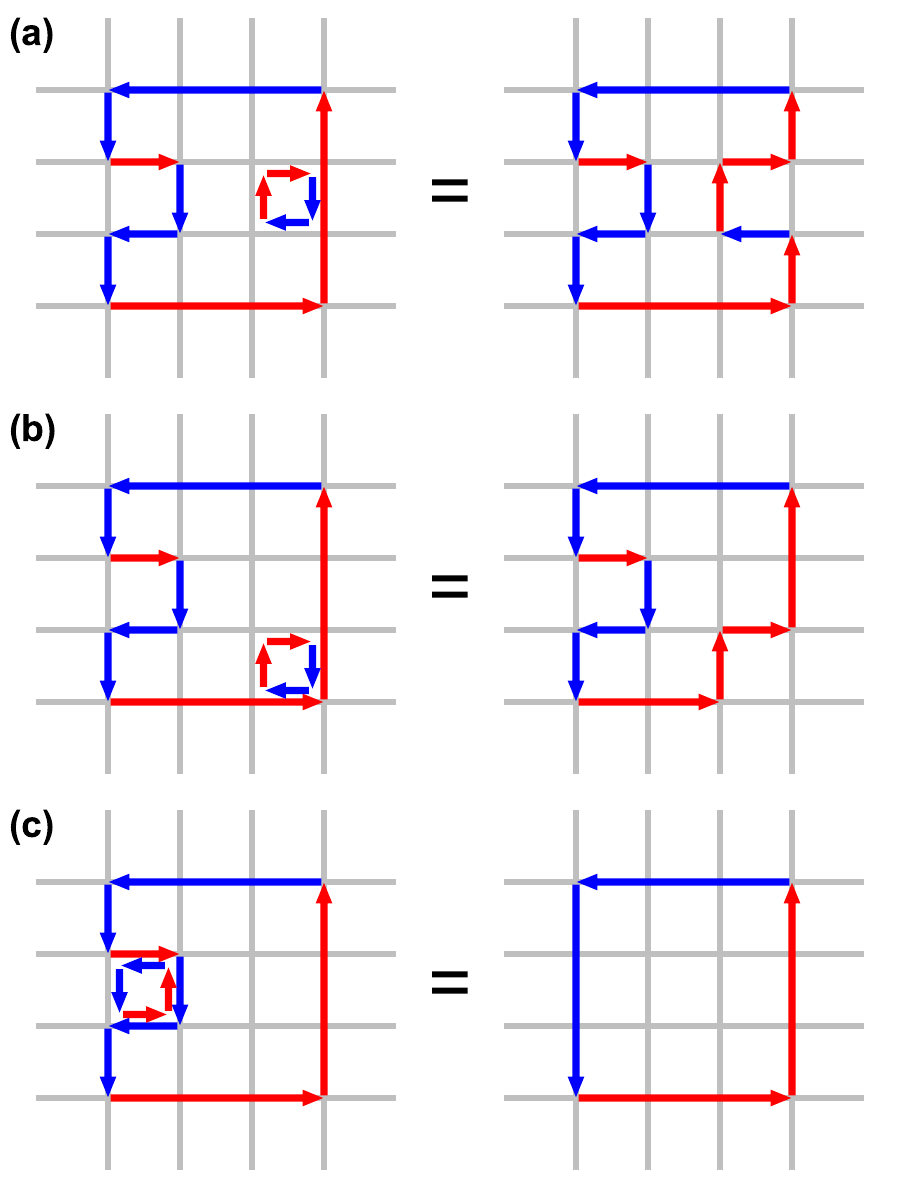}
    \caption{Depictions of the different ways in which the addition of a 
             current loop can affect the non-sparsity objective $f_S$ of 
             the current distribution in a wireframe:
             (a) a net increase of two active segments, corresponding to 
                 an increase of $f_S$ by 1;
             (b) no change in the number of active segments or $f_S$;
             (c) a net decrease of two active segments, corresponding to 
                 a decrease of $f_S$ by 1.} 
    \label{fig:current_loop_f_S}
    \end{center}
\end{figure}

It should be noted that adding a loop of current to the wireframe will not 
necessarily increase the value of $f_S$. The impact of a current loop on the
sparsity of a solution depends on the current distribution in the wireframe
before the loop is added. The possible effects of adding a current loop
are illustrated in Fig.~\ref{fig:current_loop_f_S}, which depicts the addition
of a current loop in three different locations around an initial current
distribution. In Fig.~\ref{fig:current_loop_f_S}a, in which the current loop
is added next to a straight section, the result is to cancel the current in
the segment that was previously within the straight section and add current
to the three segments that were initially inactive. This leads to a net increase
of two active segments, or an increase of $f_S$ by 1. 
In Fig.~\ref{fig:current_loop_f_S}b, a current loop is added next to a corner
of the initial current path. In this case, two initially active segments are
cancelled and two initially inactive segments are made active, resulting in
no net change in $f_S$. Finally, in Fig.~\ref{fig:current_loop_f_S}c, a
current loop is added to the inside of a c-shaped ``curve'' in the current
path. This addition cancels the currents in three initially active segments
and imparts current in one initially inactive segment, resulting in a net
reduction of two active segments, or a decrease of $f_S$ by 1. 

As an aside, note that in each case shown in Fig.~\ref{fig:current_loop_f_S}, 
the polarity of the added current loop was chosen such that there would
be no forked current paths in the final current distributions. Had the 
polarities of the loops been reversed in any of the three cases, the final
current distributions would have exhibited segments with twice the initial
current and forks in the current paths. Such cases could be prevented, for
example, by the ``no crossing'' eligibility rule described above and in 
Table \ref{tab:eligibility}.

The basic GSCO procedure is summarized formally in Algorithm~\ref{alg:gsco}.
The required inputs include $\mathbf{A}$ and $\mathbf{b}$ (Eq.~\ref{eqn:f_B})
for the computation of the objective function as well as $\mathbf{C}$ and 
$\mathbf{d}$ (Eq.~\ref{eqn:constr_all}) to evaluate the constraints on the 
segments for determining eligibility. An additional input is a set of vectors
$\{\mathbf{u} | 1 \leq i \leq \text{\# cells in wireframe} \}$, each of which
encodes a unit current in four segments around a given wireframe cell. Thus,
if a wireframe initially has a current distribution represented by $\mathbf{x}$,
the modified current distribution obtained by adding a loop of current 
$I_\text{loop}$ to the $i^{th}$ wireframe cell can be expressed as 
$\mathbf{x} \pm I_\text{loop}\mathbf{u}_i$, where the sign 
specifies the polarity of the loop. With this background data in place,
the outcome of GSCO is shaped by the initial current distribution 
$\mathbf{x}_\text{init}$, the loop current $I_\text{loop}$, the sparsity
weighting parameter $\lambda_S$, and the set of eligibility rules chosen
from Table \ref{tab:eligibility} (denoted as \textsc{Rules}) determining where 
loops may be added.

With the inputs and hyperparameters defined, the algorithm proceeds to add 
loops of current iteratively to the solution. Each iteration consists of
two steps. The first is to build up a set $L$ of 
possible current loops to add to the solution $\mathbf{x}$. This entails 
checking, for each wireframe cell, whether a loop of current
$\pm I_\text{loop}$ in 
that cell is eligible to be added according the rules.
If a loop with current $I_\text{loop}$ is eligible to be added to the
$i^\text{th}$ cell, for example, then the vector 
$I_\text{loop}\mathbf{u}_\text{i}$ is added to $L$. 
After $L$ is populated, the second step is to choose the optimal member 
$\mathbf{y}^* \in L$ that
achieves the lowest value of the optimization objective $f_\text{GSCO}$ when
added to the present solution. The solution $\mathbf{x}$ is then updated to 
$\mathbf{x} + \mathbf{y}^*$. The iterations continue until a stopping condition 
is met; namely, that no eligible cells exist or $f_\text{GSCO}$ stops decreasing
as described above.
Note that the set $L$ of eligible current loops must be reconstructed after
every iteration, as some rules depend on the current distrubtion encoded by 
$\mathbf{x}$ and thus a cell that was eligible in one
iteration will not necessarily remain eligible in a subsequent iteration.

\begin{figure}
\begin{algorithm}[H]
    \DontPrintSemicolon
    \newcommand\mycommfont[1]{\textcolor{gray}{#1}}
    \SetCommentSty{mycommfont}
    \SetKwComment{Comment}{// }{}
    \KwIn{$\mathbf{A}$, $\mathbf{b}$, $\mathbf{C}$, $\mathbf{d}$, 
          $\{\mathbf{u}_i | 1 \leq i \leq \text{\# cells in wireframe}\}$} 
    \KwOut{Optimized wireframe currents ($\mathbf{x}$)}
    \SetKwBlock{Function}{function}{end function}
    \Function($\text{GSCO} {(} \mathbf{x}_\text{init}, I_\text{loop}, 
              \lambda_S, \textsc{Rules} {)}$)
    {
        $\mathbf{x} = \mathbf{x}_\text{init}$ \;
        \Repeat{no eligible cells exist or $f_\text{GSCO}$ stops decreasing}{
            $L =\{\}$ \; 
            \For{$i = 1, ..., $ \# cells in wireframe}{
                \If{$\mathbf{x} + I_\text{loop}\mathbf{u}_i$ 
                    obeys all \textsc{Rules}} {
                    $L = L \cup \{I_\text{loop}\mathbf{u}_i\}$ \;
                }
                \If{$\mathbf{x} - I_\text{loop}\mathbf{u}_i$ 
                    obeys all \textsc{Rules}} {
                    $L = L \cup \{-I_\text{loop}\mathbf{u}_i\}$ \;
                }
            }
            $\mathbf{y}^* = 
                \min_{\mathbf{y} \in L} 
                \left[ f_B(\mathbf{x} + \mathbf{y}) 
                    + \lambda_S f_S(\mathbf{x} + \mathbf{y}) \right]$ \;
            $\mathbf{x} = \mathbf{x} + \mathbf{y}^*$ \;
        }
        \Return{$\mathbf{x}$} \;
    }
    \caption{Greedy Stellarator Coil Optimization (GSCO)}
    \label{alg:gsco}
\end{algorithm}
\end{figure}

Throughout this section, ``coils'' will be used as shorthand to describe 
isolated, filamentary paths of current flowing through the wireframe. 
These paths, which contain many sharp, angular turns, would not 
necessarily constitute feasible coil geometry -- especially if it is desired
to make the coils with high-temperature superconducting tape, which cannot
tolerate sharp bends \cite{vanderlaan2019a, paz-soldan2020a}. 
However, they could form a
starting point for more realistic coil designs. For example, a wireframe
current path with sharp angles could be converted to a smoother curve using
splines. Furthermore, these smoothed coil shapes could be refined 
for improved field accuracy and other objectives with an optimizer such as
\textsc{Focus} \cite{zhu2018a}, \textsc{Desc} \cite{dudt2020a}, 
\textsc{Gospel} \cite{yamaguchi2021a}, or 
\textsc{Simsopt} \cite{wechsung2022a}.

The GSCO procedure can design coils in many different styles and topologies
depending on how it is initialized and constrained. In the following sections, 
some example solutions will be presented to demonstrate some of the 
possibilities.

\subsection{Modular coils}
\label{sec:gsco_modular}

The GSCO algorithm will first be applied to the design of modular coils for 
the Precise QA equilibrium. Modular coils, which encircle the plasma once
poloidally while remaining toroidally localized, are employed in most modern 
stellarator designs. 

It is important to note that the GSCO algorithm as formulated in 
Sec.~\ref{sec:gsco_method} cannot itself create new modular coils on an 
empty toroidal wireframe grid. Since the incremental current loops added during
each iteration cannot, by construction, encircle the plasma, no sum of these
loops can contribute a net poloidal current. However, if the wireframe is 
initialized with some poloidal current flows, GSCO can reshape those flows to 
create an accurate magnetic field. 

Fortunately, the initial poloidal flows need not be complicated. In particular,
they may take the form of simple poloidally-flowing rings within the wireframe.
The only requirement is that the sum of the currents in these initial flows
must equal the net poloidal current corresponding to the desired average
toroidal magnetic field on axis.

\begin{figure*}
    \begin{center}
    \includegraphics[width=\textwidth]{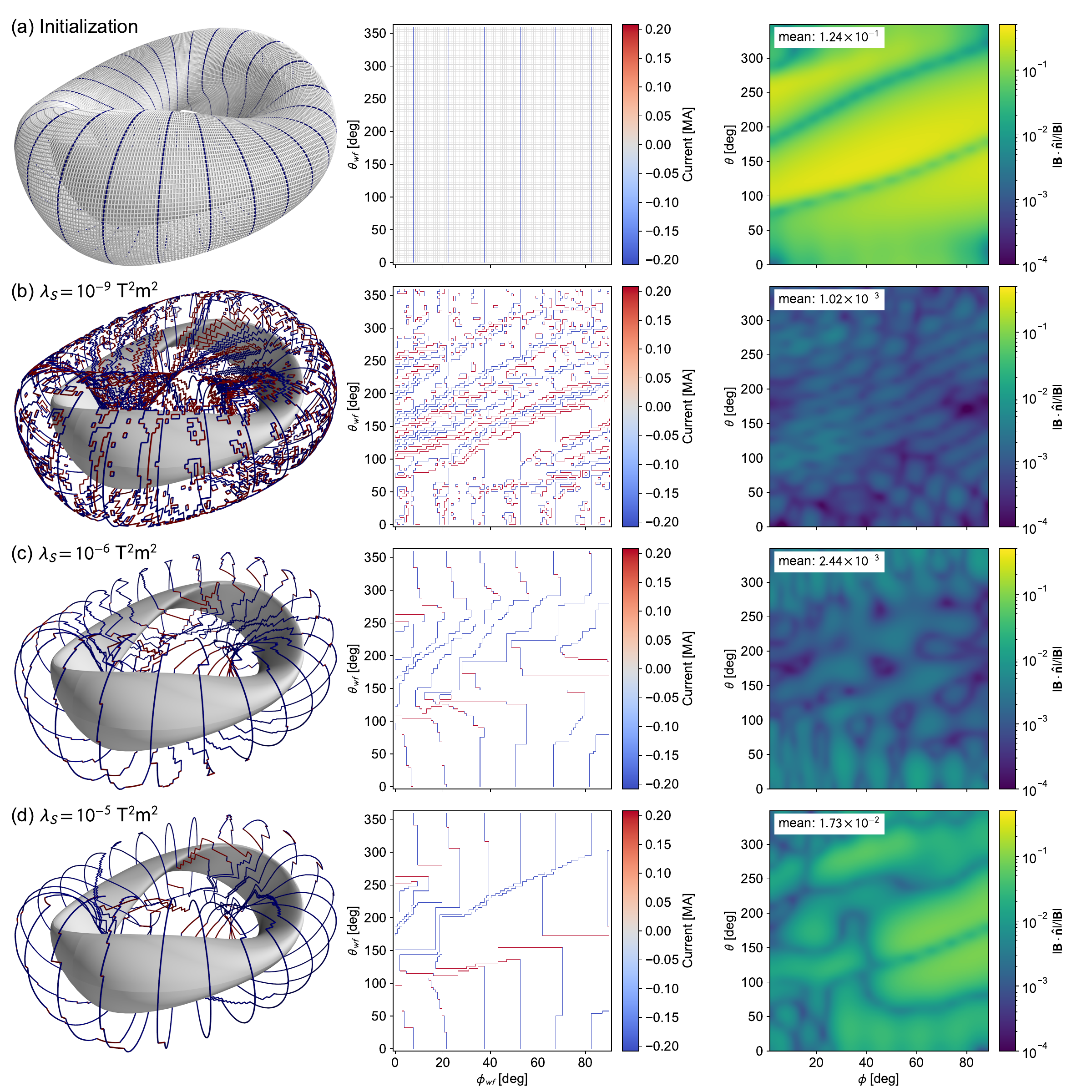}
    \caption{Current distributions (left and center column) and relative 
             normal magnetic fields on the target plasma boundary (right
             column) for 
             (a) a wireframe initialized with planar coils,
             and GSCO solutions using
             (b) $\lambda_S = 10^{-9}$ T$^2$m$^2$,
             (c) $\lambda_S = 10^{-6}$ T$^2$m$^2$, and
             (d) $\lambda_S = 10^{-5}$ T$^2$m$^2$.
             In subplots (b)-(d), inactive wireframe segments are hidden.
             The plots in the middle and right columns each represent one 
             half-period of the wireframe and the plasma boundary, 
             respectively.}
    \label{fig:gsco_modular_solns}
    \end{center}
\end{figure*}

One example initialization is shown in Fig.~\ref{fig:gsco_modular_solns}a.
The wireframe used in this case has a grid resolution of $96 \times 100$,
which is much greater than the resolutions used for the RCLS solutions in
Sec.~\ref{sec:rcls}. The higher resolution used for the GSCO solutions in this
section compensates for the much more stringent limitations that the GSCO
solver effectively places on the currents that can be carried by each segment.
The wireframe is initialized to have six planar poloidal loops of current per 
half-period, or twenty-four poloidal loops altogether. The initial loops are
spaced evenly in the toroidal angle. The segments in the poloidal current 
loops each carry 0.208~MA, adding to a total of 5~MA of poloidal current.

A series of optimizations was performed, using GSCO from this initialization
of the wireframe, with different values of the weighting parameter $\lambda_S$.
In each optimization, the eligibility rule preventing crossing current paths
was in effect. In addition, the current used in each loop to be added to the 
wireframe matched the magnitude of the current in the initialized planar
poloidal coils. This way, loops could be added next to those initialized 
coils to modify their shapes without creating forked current paths, similarly
to the example in Fig.~\ref{fig:loop_schematic}b.
Current loops were also eligible to be added to inactive segments in the 
spaces between existing coils, thereby creating new saddle coils.

\begin{figure}
    \begin{center}
    \includegraphics[width=0.5\textwidth]{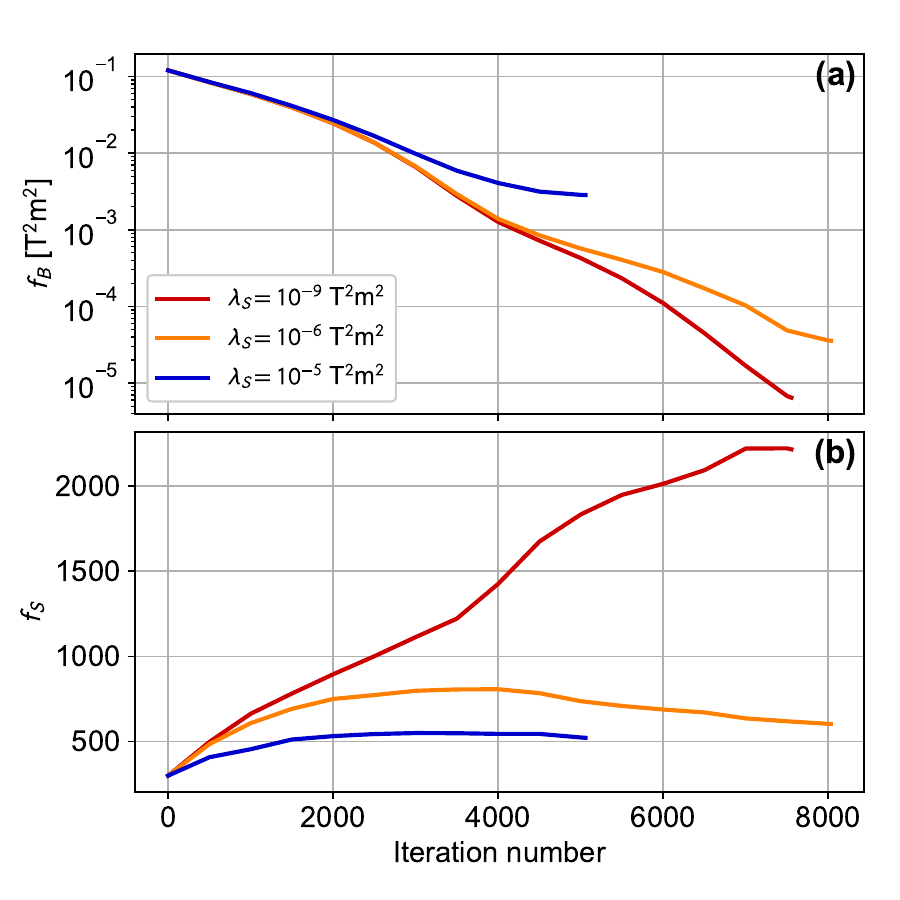}
    \caption{Values of sub-objective functions over the course of the
             iterations for GSCO performed on a modular coil initialization
             with different values of the weighting factor $\lambda_S$.
             (a) $f_B$; (b) $f_S$.}
    \label{fig:gsco_modular_iterations}
    \end{center}
\end{figure}

The trends in the sub-objective functions over the course of the optimizations
for three selected values of $\lambda_S$ are shown in 
Fig.~\ref{fig:gsco_modular_iterations}. In every case, the optimizations 
terminated upon reaching a minimal value of the total objective function 
$f_{GSCO}$. As expected, lower values of 
$\lambda_S$ tend to result in solutions with higher field accuracy 
(lower $f_B$) and lower sparsity (higher $f_S$). Interestingly, for the 
intermediate value of $\lambda_S=10^{-6}$ T$^2$m$^2$, both sub-objectives
decrease simultaneously during roughly half of the iterations. 

The plots in Fig.~\ref{fig:gsco_modular_solns}b-d show the results
of a series of optimizations using different values of the sparsity weighting 
factor $\lambda_S$. Each optimization began from the same planar loop 
initialization shown in Fig.~\ref{fig:gsco_modular_solns}a. The solutions
differ starkly, both in terms of the qualitative appearance of the current
distribution (left and center columns) and in terms of the field accuracy
(right column). 

For the lowest value of $\lambda_S$ ($10^{-9}$ T$^2$m$^2$,
Fig.~\ref{fig:gsco_modular_solns}b), the current distribution has a highly 
intricate appearance. The inboard side of the wireframe is nearly filled with
active segments, forming a dense network of current paths consisting of both
modular coils that encircle the plasma and saddle coils that do not. The 
outboard side is less dense with current-carrying segments but still features
areas packed densely with saddle coils. The solution includes a number of
miniature saddle coils that enclose just one or two cells of the wireframe.
On the other hand, the solution achieves the best field accuracy metrics,
with a surface-averaged relative normal field on the target plasma boundary of
$\mrbn = 1.02 \times 10^{-3}$.

For the intermediate value of $\lambda_S$ ($10^{-6}$ T$^2$m$^2$,
Fig.~\ref{fig:gsco_modular_solns}c), the current distribution is much more
sparse. It consists almost purely of six modular coils per
half period, with the exception of one small saddle coil near 
$\phi_\text{wf}=20^\circ$ and $\theta_\text{wf}=140^\circ$.
Thus, in this case, the predominant effect of GSCO
was to reshape the planar coils from the initialization rather than to add lots
of new saddle coils. Overall, the solution has 72\% fewer active segments than
the solution with $\lambda_S=10^{-9}$ T$^2$m$^2$. The increase in sparsity and
simplicity came at the expense of some field accuracy, with $\mrbn$ increasing
to $2.44 \times 10^{-3}$.

For the highest value of $\lambda_S$ ($10^{-5}$ T$^2$m$^2$,
Fig.~\ref{fig:gsco_modular_solns}d), the current distribution is even more
sparse, and this time is a true modular coil solution with no saddle coils
present. However, $\mrbn$ increased substantially relative to the previous
case, to $1.73 \times 10^{-2}$.

In the solutions using $\lambda_S = 10^{-6}$ and $10^{-5}$ T$^2$m$^2$, it is
notable that many of the initialized planar coils underwent little to no
geometric modification on the outboard side (see, in particular, 
Fig.~\ref{fig:gsco_modular_solns}c-d where $\theta_{wf} < 100^\circ$ and 
$\theta_{wf} > 300^\circ$). This is 
reminiscent of a pilot plant design described in 
Refs.~\cite{brown2015a,gates2017a}. For that design, the \textsc{Coilopt++}
code was employed to optimize the geometry of the modular coils, modeled
as spline curves with constraints applied to be mostly vertical on the outboard
side. In the GSCO solutions shown here, by contrast,
no specific constraints were applied to keep the coils vertical on the
outboard side; rather, this geometry is an indirect result of the application
of the sparsity objective and the natural tendency for more shaping to be
required on the inboard side of most stellarator designs.

The optimizations shown in Fig.~\ref{fig:gsco_modular_solns}c-d were performed
using 64 cores on a 2.9 GHz Intel Cascade Lake processor node on the Stellar
cluster at Princeton University \cite{stellar_cluster}. 
The wall clock times ranged from 44 seconds
($\lambda_S = 10^{-9}$ T$^2$m$^2$) to 75 seconds 
($\lambda_S = 10^{-6}$ T$^2$m$^2$), corresponding to 0.8 to 1.3 CPU-hours.
In each case, the wireframe contained a total of 19,200 segments per
half-period, and 1,024 field
evaluation points were used in the target plasma boundary to evaluate $f_B$.

Before proceeding with further optimization results, it is worthwhile to 
note the extent to which the geometry of the wireframe, which remains fixed
during each optimization, can affect the quality of the solutions. In some
cases, even subtle changes in the wireframe geometry can have a 
noticeable impact on the achievable flux surface quality. This is discussed
in more detail in Appendix \ref{apx:wframe_geometry}.

\subsection{Sector-confined saddle coils}
\label{sec:gsco_sector}

While the modular coil optimizations in Sec.~\ref{sec:gsco_modular} serve as a
useful proof-of-concept for the GSCO procedure, they do not utilize a key
feature of GSCO; namely, the ability to impose arbitrary spatial restrictions 
on where coils may appear. Such restrictions enable the development of designs
that explicitly reserve space for other device components or facilitate
easier assembly and disassembly.

\begin{figure}
    \begin{center}
    \includegraphics[width=0.5\textwidth]{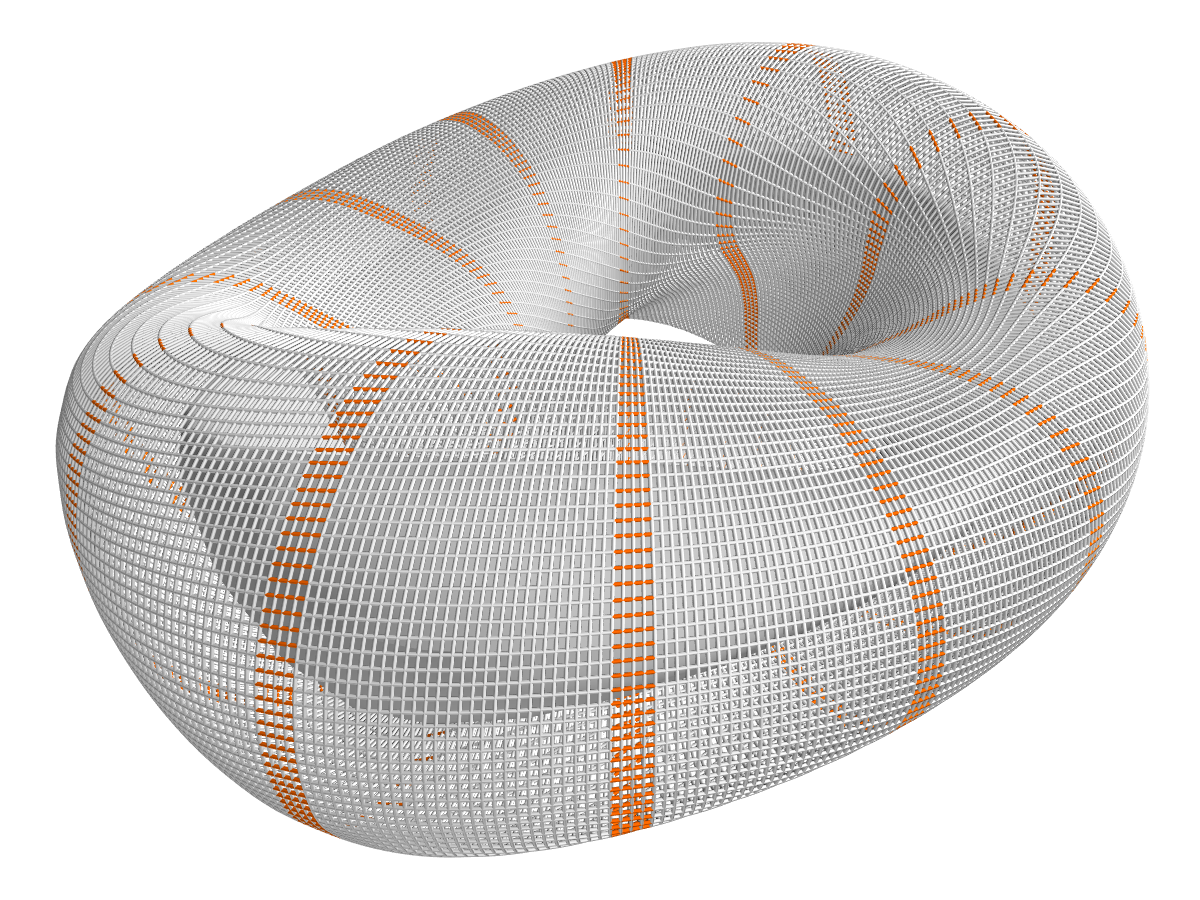}
    \caption{Depiction of a wireframe with sector constraints that prevent
             current from crossing certain planes of constant toroidal angle.
             The constrained segments, through which no current may flow,
             are shown in orange.}
    \label{fig:sector_constraints}
    \end{center}
\end{figure}

One potential design concept that would be simple to assemble is one in which
all coils are constrained to lie within toroidal sectors; i.e. in which there
are designated planes of constant toroidal angle that no coils cross. To 
attain a design like this with GSCO, the first step is to impose constraints
on a wireframe that prevent current from crossing certain toroidal planes.
Such a set of constraints is displayed in Fig.~\ref{fig:sector_constraints}.
In this schematic, the orange segments are subject to the segment constraints
of Eq.~\ref{eqn:constr_segment}. Hence, they may not carry any current and,
for the purposes of GSCO, a cell that contains one or more of these constrained
segments is not eligible for the addition of a loop of current. Note that the
constraints here are only applied to \textit{toroidal} segments; hence, it is 
allowable to add planar coils within these constrained regions in which 
the current flows in a strictly \textit{poloidal} direction.

\begin{figure}
    \begin{center}
    \includegraphics[width=0.45\textwidth]{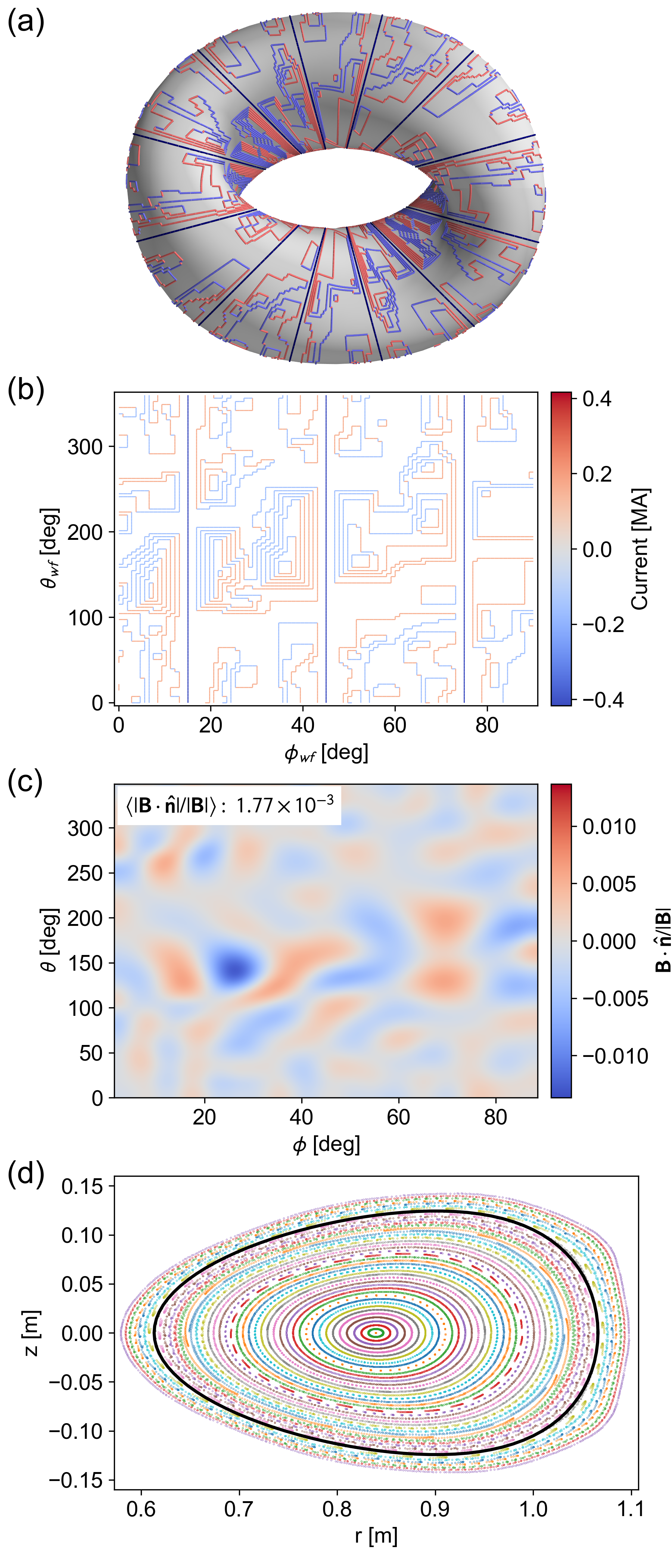}
    \caption{Current distribution and magnetic field properties of a wireframe
             containing a set of planar TF coils and saddle coils produced
             by GSCO constrained to lie within toroidal sectors.
             (a) 3D rendering of the active segments superimposed, for visual
                 clarity, on a torus that fills the volume of the wireframe;
             (b) 2D rendering of the currents in one half-period of the 
                 wireframe;
             (c) relative normal component of the magnetic field on one 
                 half-period of the target plasma boundary;
             (d) Poincar\'e cross-section of field lines produced by the
                 wireframe distribution, with the target plasma boundary
                 shown as the black curve.}
    \label{fig:gsco_sector}
    \end{center}
\end{figure}

A sample solution that uses these constraints is shown in 
Fig.~\ref{fig:gsco_sector}. The wireframe was initialized with a set of 
planar, poloidal current flows to produce the required toroidal field. These 
current flows were placed inside the regions where the poloidal segments
were constrained and loops of current could not be added during GSCO; hence, 
their shapes would remain unchanged. There are only three planar coils per
half-period in this example, in contrast to the six used in the modular
coil solution in Sec.~\ref{sec:gsco_modular}, although the net poloidal current
is the same (5 MA), producing an average field on axis of 1 T.

For the optimization in Sec.~\ref{sec:gsco_modular}, the loop current in 
GSCO needed to equal the current in the initialized planar coils so that 
their shapes could be modified by current loops without creating forked current
paths. In this case, by contrast, the optimization will not place current loops
adjacent to the planar coils. Thus, the choice of loop current is arbitrary. 
To determine which current level was best suited for this application, 
multiple optimizations were performed with different loop current levels.
In the optimizations performed to date, the best results have been achieved for 
this configuration with a loop current of 0.15 MA, or 3\% of the total poloidal 
current.

The solution shown in Fig.~\ref{fig:gsco_sector} was run with a loop current
of 0.15 MA and $\lambda_S=10^{-7.5}$ T$^2$m$^2$. Renderings of the active 
segments of the final current distribution are rendered in 3D in 
Fig.~\ref{fig:gsco_sector}a and in 2D (for one half-period) in 
Fig.~\ref{fig:gsco_sector}b. For visual clarity, the wireframe segments in
the 3D rendering are superimposed on an opaque toroidal surface. The 3D
rendering is viewed from above, such that the initialized planar coils
appear as straight lines radiating from the center. In the 2D rendering,
the planar coils appear as straight, vertical lines. The field created by
this current distribution is highly accurate, with $\rbn$ rarely exceeding
1\% (Fig.~\ref{fig:gsco_sector}c) and with flux surfaces that exhibit excellent
agreement with the target plasma boundary (Fig.~\ref{fig:gsco_sector}d).

As a result of the constraints placed on the optimization, the
saddle coils are conveniently confined to sectors between the planar coils.
With this layout, the device would likely be easier and faster to assemble and 
disassemble for maintenance. From the standpoint of a power plant, any 
reduction of maintenance time would make the plant more economical. 

The current distribution in Fig.~\ref{fig:gsco_sector}a-b is substantially less 
sparse than the modular coil solutions in Figs.~\ref{fig:gsco_modular_solns}c-d.
However, note that the current paths in this solution tend to appear in 
concentric loops with like polarities, particularly on the inboard side 
(Fig.~\ref{fig:gsco_sector}b, $100^\circ < \theta_{wf} < 260^\circ$). Thus, 
rather than interpreting each of these concentric loops as an individual coil, 
it may be more useful to view the sets of concentric 
loops as single saddle coils with finite dimensions, with the individual 
filaments forming the basis for the design of a winding pack.

\subsection{Multiple currents}
\label{sec:gsco_multicur}

\begin{figure}
    \begin{center}
    \includegraphics[width=0.45\textwidth]{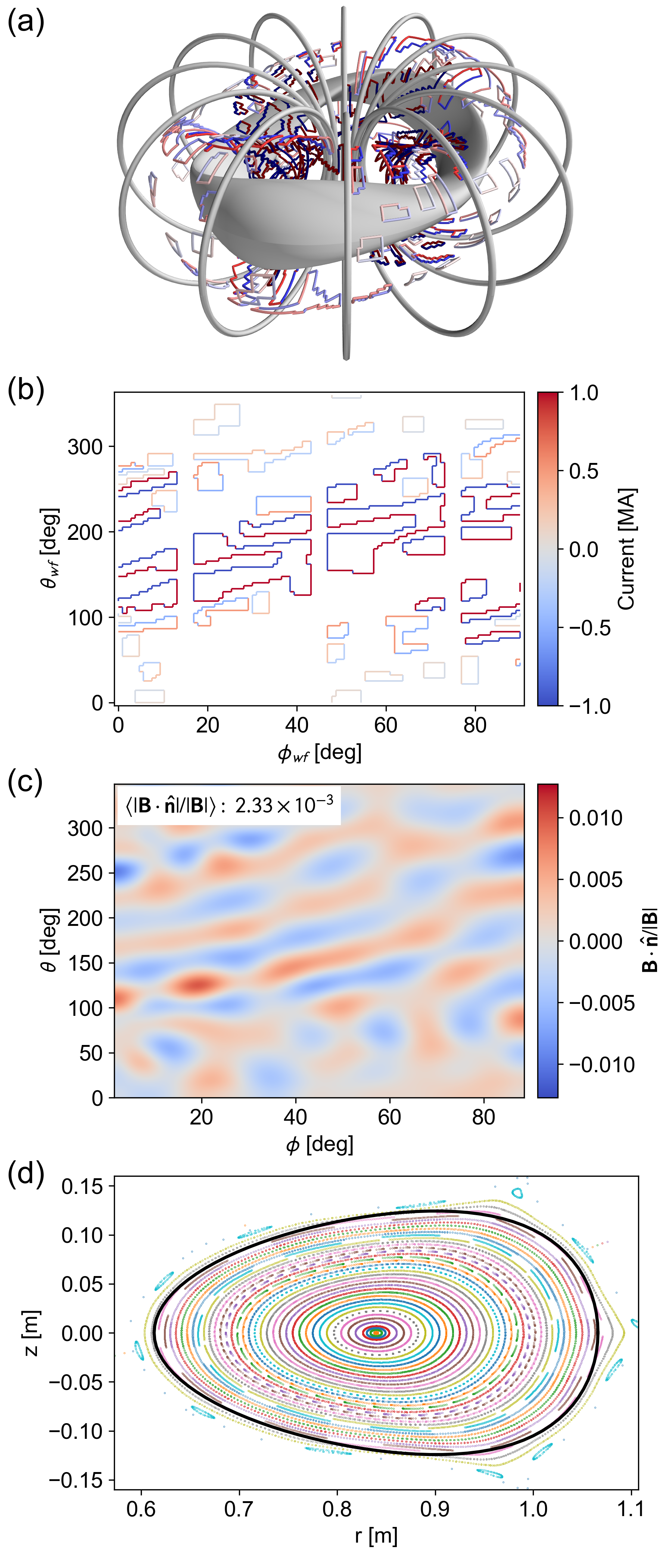}
    \caption{Current distribution and magnetic field properties of a wireframe
             with saddle coils designed in a sequence of GSCO procedures.
             (a) 3D rendering of the active segments along with accompanying
                 circular TF coils and the target plasma boundary;
             (b) 2D rendering of the currents in one half-period of the 
                 wireframe;
             (c) relative normal component of the magnetic field on one 
                 half-period of the target plasma boundary;
             (d) Poincar\'e cross-section of field lines produced by the
                 wireframe distribution, with the target plasma boundary
                 shown as the black curve.}
    \label{fig:gsco_seq}
    \end{center}
\end{figure}

The solutions presented in the previous sections each utilized a single run 
of GSCO with fixed values of hyperparameters such as the loop current 
$I_\text{loop}$ and weighting factor $\lambda_S$. One consequence of this was 
that all coils 
constucted by the GSCO algorithm carried the same current. However, it could
be advantageous seek solutions in which different coils carry different 
currents. Such solutions can be attained by performing a sequence of GSCO 
procedures with varying parameters at each step. 
One example is shown in Fig.~\ref{fig:gsco_seq}, and the associated procedure
is summarized in Algorithm \ref{alg:gsco_multi}.

\begin{figure}
\begin{algorithm}[H]
    \DontPrintSemicolon
    \KwIn{$\mathbf{A}$, $\mathbf{b}$, $\mathbf{C}$, $\mathbf{d}$, 
          $\{\mathbf{u}_i | 1 \leq i \leq \text{\# cells in wireframe}\}$}
    \KwOut{Optimized wireframe currents ($\mathbf{x}$)}
    \SetKwBlock{Function}{function}{end function}
    \Function($\textsc{GSCO\_multistage} {(} \mathbf{x}_\text{init}, 
              I_\text{start}, \lambda_S, N_\text{cells,min} {)}$) {
        \textsc{Rules1} = \{wireframe constraints, no crossing, 
                            max loops per cell (1)\} \;
        $\mathbf{x} = \mathbf{x}_\text{init}$ \;
        $I_\text{loop} = I_\text{start}$ \;
        \Repeat{$\mathbf{x}$ does not change from previous iteration} {
            $\mathbf{x} = \text{GSCO}(\mathbf{x}, I_\text{loop}, \lambda_S, 
                \textsc{Rules1})$ \;
            Remove coils from $\mathbf{x}$ enclosing fewer cells than 
                $N_\text{cells,min}$ \;
            Update $\textbf{C}$ and $\textbf{d}$ to constrain segments within 
                new coils to carry no current \;
            $I_\text{loop} = I_\text{loop}/2$ \;
        }
        Update $\textbf{C}$ and $\textbf{d}$ to remove constraints on segments 
            within coils \;
        $\textsc{Rules2} = \textsc{Rules1} \cup \{\text{no new coils}\}$ \;
        $\mathbf{x} = \textsc{GSCO\_match}(\mathbf{x}, 0, \lambda_S, 
            \textsc{Rules2})$ \;
        \Return{$\mathbf{x}$}
    }
    \caption{Multistage GSCO}
    \label{alg:gsco_multi}
\end{algorithm}
\end{figure}

The procedure applied the standard GSCO function (Algorithm \ref{alg:gsco}) 
multiple times, each time using a different loop current $I_\text{loop}$.
In this way, the solution obtained a few new coils at each step with a 
distinct current level. The sequence proceeded in a ``drop-down''
manner, with the first optimization using the highest current 
($I_\text{loop} = $ 1 MA), and with each subsequent optimization using 
one-half of the loop current of the previous optimization. 

The procedure also took measures to avoid potentially inconvenient features 
such as concentric coils which had been prevalent in the solution in 
Sec.~\ref{sec:gsco_sector}, or very small coils. To lessen the likelihood
of concentric coils appearing, the each GSCO optimization was run with the 
``max loops per cell'' rule (Table \ref{tab:eligibility}) with $N_\text{max}=1$.
In addition, following each GSCO step, all segments enclosed within existing 
coils were constrained to carry zero current.
To avoid the presence of overly small saddle coils, any coil at the end of an 
optimization that enclosed fewer than 20 cells was eliminated before starting 
the subsequent optimization. Presumably, if a very small coil appears
in a given region during one optimization in the sequence, a larger coil
would be placed in the same location with half the current in the next
optimization.

\begin{figure}
\begin{algorithm}[H]
    \DontPrintSemicolon
    \newcommand\mycommfont[1]{\textcolor{gray}{#1}}
    \SetCommentSty{mycommfont}
    \SetKwComment{Comment}{// }{}
    \KwIn{$\mathbf{A}$, $\mathbf{b}$, $\mathbf{C}$, $\mathbf{d}$, 
          $\{\mathbf{u}_i | 1 \leq i \leq \text{\# cells in wireframe}\}$} 
    \KwOut{Optimized wireframe currents ($\mathbf{x}$)}
    \SetKwBlock{Function}{function}{end function}
    \Function($\textsc{GSCO\_match} {(} \mathbf{x}_\text{init}, 
              I_\text{default}, \lambda_S, \textsc{Rules} {)}$)
    {
        $\mathbf{x} = \mathbf{x}_\text{init}$ \;
        \Repeat{no eligible cells exist or $f_\text{GSCO}$ stops decreasing}{
            $L =\{\}$ \; 
            \For{$i = 1, ...,  \text{\# cells in wireframe}$} {
                $\{I_j | 1 \leq j \leq 4\} =$ the currents in the segments
                    around cell $i$ \;
                \If {$I_j = 0$ for all $j$} {
                    $I_\text{loop} = I_\text{default}$
                }
                \ElseIf {the nonzero $I_j$ have a single absolute value 
                         ($\equiv I_\text{match}$)} {
                    $I_\text{loop} = I_\text{match}$ \;
                }
                \Else {
                    Not eligible; continue to next cell
                }
                \If{$\mathbf{x} + I_\text{loop}\mathbf{u}_i$ 
                    obeys all \textsc{Rules}} {
                    $L = L \cup \{I_\text{loop}\mathbf{u}_i\}$ \;
                }
                \If{$\mathbf{x} - I_\text{loop}\mathbf{u}_i$ 
                    obeys all \textsc{Rules}} {
                    $L = L \cup \{-I_\text{loop}\mathbf{u}_i\}$ \;
                }
            }
            $\mathbf{y}^* = 
                \min_{\mathbf{y} \in L} 
                \left[ f_B(\mathbf{x} + \mathbf{y}) 
                    + \lambda_S f_S(\mathbf{x} + \mathbf{y}) \right]$ \;
            $\mathbf{x} = \mathbf{x} + \mathbf{y}^*$ \;
        }
        \Return{$\mathbf{x}$} \;
    }
    \caption{GSCO with current matching}
    \label{alg:gsco_match}
\end{algorithm}
\end{figure}

After several sequential GSCO procedures with decreasing loop current levels,
a final GSCO procedure was performed to fine-tune the shapes of the existing 
coils without adding any new coils. Practically, this was accomplished by
removing the constraints on segments enclosed by coils and
invoking the ``no new coils'' rule (Table \ref{tab:eligibility}) with
an enhanced version of the original GSCO function 
that offers the option to match existing currents in the wireframe.
This is summarized in Algorithm \ref{alg:gsco_match} as the
function \textsc{GSCO\_match}. The current-matching feature is implemented
in each iteration through an additional step performed prior
to checking the eligibility of a given cell. Specifically, if the 
cell contains segments that are already part of an existing coil with a current 
$I_\text{match}$, the optimizer will consider a loop current for that cell
with a current equal to $I_\text{match}$ rather than the default current
$I_\text{default}$ supplied by the user as a hyperparameter. This is 
advantageous when the initial
current distribution $\text{x}_\text{init}$ contains coils with different 
currents. When the greedy optimizer has the flexibility to add loops of current
that match the current of the coils that the loops are next to, the optimzer
is able to adjust the shapes of all the coils in a single function call.

Note that if a single cell contains segments utilized by different coils 
with different current levels, it is also considered ineligible for the 
addition of a current loop, as the value to use for $I_\text{match}$ is 
ambiguous. Furthermore, adding a loop with any (nozero) value of 
$I_\text{match}$ in this case would create a forked current path. 

In general, if a given cell has no current-carrying segments (i.e., the cell
is not adjacent to an existing coil), \textsc{GSCO\_match} can consider
adding a loop of current with a user-specified default current level.
However, for the application in the multistage GSCO procedure
(Algorithm \ref{alg:gsco_multi}), this feature was circumvented by
applying the ``no new coils'' rule and (redundantly) setting the 
$I_\text{default}$ hyperparameter to 0. This is done because the goal at
this final step of Algorithm \ref{alg:gsco_multi} is strictly to refine the
shapes of the existing coils rather than to add any new ones.

Applying the multistage procedure (Algorithm \ref{alg:gsco_multi}) to a 
wireframe around the Precise QA equilibrium yielded the solution shown
in Fig.~\ref{fig:gsco_seq}a-b. 
Similarly to solution in Sec.~\ref{sec:gsco_sector}, the saddle coils added
by the GSCO procedures are confined to toroidal sectors. One new
aspect of this solution is that the toroidal field is produced not by the 
wireframe itself but by a set of external, circular 
TF (toroidal field) coils. Therefore, the wireframe was not initialized
with any poloidal current flows. Hence, this solution demonstrates how
a wireframe can be optimized to improve the field created by a pre-existing set
of coils.

The multistage procedure used $\lambda_S = 1 \times 10^{-7}$ T$^2$m$^2$. 
The starting loop current $I_\text{start}$ was 1 MA. Ultimately, the 
successive applications of GSCO added coils with currents of
1 MA, 500 kA, 250 kA, 125 kA, and 62.5 kA; subsequent GSCO runs with lower
currents failed to add any coils above the minimum size $N_\text{cells,min}$. 
The computation time to perform Algorithm \ref{alg:gsco_multi} for this solution
was substantially lower than the time required to perform 
Algorithm \ref{alg:gsco} for the solutions in Sec.~\ref{sec:gsco_modular}
and \ref{sec:gsco_sector}. This is due to the more stringent constraints
used in each optimization, which greatly reduced the number of eligible 
cells for the optimizer to check during each iteration. Using 4 cores on a 
2.9 GHz Intel Cascade Lake processor node on the Stellar cluster, the longest 
call to GSCO in the sequence took 34 seconds, or 136 CPU-seconds. 
Running the entire procedure
on a laptop, including additional steps for setting up the wireframe and 
post-processing, took less than 5 minutes.

Remarkably, the solution leaves a substantial amount of empty space on the
outboard side, particularly for $\theta_{wf} < 100^\circ$ and 
$15^\circ < \phi_{wf} < 45^\circ$ in Fig.~\ref{fig:gsco_seq}b. 
This is the case despite
no specific spatial restrictions having been placed on this area. 
Nevertheless, the field accuracy is quite good, with $\rbn$ is mostly below
1\% (\ref{fig:gsco_seq}c). The field exhibits decent flux surface agreement 
(Fig.~\ref{fig:gsco_seq}d), although the flux surfaces near the plasma 
boundary appear to be deformed slightly by a resonant perturbation just 
outside. Field accuracy could be improved by running the
optimization with a lower value of $\lambda_S$, although this would come at 
the expense of adding more coils with more complicated geometry.


\section{Summary and conclusions}
\label{sec:conclusions}

In this paper, the wireframe framework for stellarator coil design and 
optimization was introduced. Consisting of a set of interconnected 
current-carrying segments, the wireframe offers a solution space with a
spatially local parametrizaton that enables the imposition of arbitrary
constraints on where current flows may be located. Two methods of optimizing
the wireframe segment currents were introduced: Regularized Constrained
Least Squares (RCLS) and Greedy Stellarator Coil Optimization (GSCO). 

RCLS is a rapid, linear optimizer that can produce highly accurate solutions 
on coarse grids, although the prevalence of junctions of segments that each
have unique current levels could make these distributions difficult to realize
in practice. 

GSCO is a fully discrete algorithm that can
construct flows of current in the wireframe that avoid the complicated 
current junctions seen in the RCLS solutions and control the number of 
unique current levels required to implement the solution. GSCO is also 
topologically flexible and can both add new coils and reshape pre-existing
coils in a single optimization. As demonstrated in Sec.~\ref{sec:gsco},
it could produce both modular coil solutions and saddle coil solutions for
the same target plasma equilibrium, with the nature of the solution dependent
on the initialization and constraints imposed.

The example solutions for both optimization methods illustrated the ease with
which arbitrary spatial constraints may be placed on current distributions.
Since the current distribution is parametrized according to the current 
carried by each individual segment, such spatial constraints can be 
implemented simply by zeroing out the parameters 
corresponding to any subset of segments where current flow is not desired. For 
example, current distributions can be constrained to leave space for ports
(Fig.~\ref{fig:wf_rcls_ports}), or to keep current flows confined to 
predefined toroidal sectors for ease of assembly (Figs.~\ref{fig:gsco_sector}
and \ref{fig:gsco_seq}).

While the current flow patterns in the GSCO solutions cannot themselves
constitute designs for coils due to their filamentary nature and the sharp
corners at the wireframe nodes, they can serve as a starting point for coil
geometry that can be subsequently refined with space curve optimization 
techniques. The value of the GSCO procedure in this case lies in its ability
to:
\begin{enumerate}
    \item identify the most important locations for saddle coils 
          and/or shape modifications for pre-existing coils, 
    \item determine coil shapes within the wireframe solution space that 
          produce a highly accurate field,
    \item obey spatial constraints that would be more difficult to enforce with 
          other optimization methods, and 
    \item exhibit flexibility in determining the number of coils.
\end{enumerate}

There are many possibilities for expanding the wireframe framework and 
associated optimization methods to increase flexibility and access larger
solution spaces. For example, with the GSCO procedure, many solutions can
only be found through optimization in multiple stages that use different
hyperparameters, as was the case for the multi-current solution in 
Sec.~\ref{sec:gsco_multicur}. By further developing customized recipes for
GSCO sequences, it should be possible to find solutions that are better
suited for a variety of applications and device requirements.

In addition, there are ways of generalizing the wireframe itself that may
lead to better solutions. For one, it may be worth experimenting with different
segment meshing structures (e.g. triangles). Furthermore,
all of the wireframes utilized in this paper exhibited two-dimensional, toroidal
topology, and were thus analogous to the toroidal winding surfaces used
in many existing coil design techniques. The use of this type of wireframe
effectively constrains the optimized current distribution to exist on a 
pre-specified surface, which puts strict limits the solution space. However,
in principle, wireframes can be constructed with three-dimensional topology,
with segments filling a volume around the plasma and thus enabling a much
more extensive solution space. Such wireframes will be developed and
investigated as next steps. 

\section{Acknowledgments}
\label{sec:acknowledgments}

The research described in this paper was conducted under the Laboratory 
Directed Research and Development (LDRD) Program at the Princeton Plasma 
Physics Laboratory, a national laboratory operated by Princeton University for 
the U.S.~Department of Energy under Prime Contract No.~DE-AC02-09CH11466. 
The United States Government retains a non-exclusive, paid-up, irrevocable, 
world-wide license to publish or reproduce the published form of this 
manuscript, or allow others to do so, for United States Government purposes.

\section{Data and code availability}
\label{sec:data_code}

The methods developed for this paper are implemented in the open-source
\textsc{Simsopt} codebase \cite{simsopt}. The three-dimensional plots were
rendered with the Mayavi scientific visualization package 
\cite{ramachandran2011a}. The data presented here, along with scripts for
reproducing the key results, are available at 
\url{https://doi.org/10.34770/6pbm-9981}.

\appendix
\section{Derivation of the formula for the magnetic field from a straight 
         wire segment}
\label{apx:segment_field_derivation}

The formula in Eq.~\ref{eqn:segment_Bfield} for the magnetic field generated by 
a straight segment of wire may be derived from the Biot-Savart law for the
magnetic field from a general space curve. The segment formula is sometimes
utilized to calculate the field from filamentary coils in which space curves
are approximated as polygons. This approach is taken, for example, by the
FLARE code \cite{frerichs2024a} and by the 
MAKEGRID code in the STELLOPT codebase \cite{stellopt}.
Derivations using alternate approaches may be found in 
Refs.~\cite{anderson1976a} and \cite{hanson2002a}.

To compute the expression for the magnetic field $\mathbf{B}$ at an evaluation 
point $\mathbf{p}_\text{o}$ from a current $I$ flowing in a segment of wire
from point $\mathbf{p}_\text{1}$ to point $\mathbf{p}_\text{2}$, one begins
with the Biot-Savart integral:

\begin{equation}
    \mathbf{B} = \frac{\mu_0}{4 \pi}
        \int_{\mathbf{p}_\text{1}}^{\mathbf{p}_\text{2}}
            \frac{I~\mathbf{dl'} \times \mathbf{r}'}{|\mathbf{r}'|^3}
\end{equation}

\noindent Here, $\mathbf{dl}'$ is a differential element of the wire segment 
and $\mathbf{r}'$ is a vector from the differential element to the field
evaluation point $\mathbf{p}_\text{f}$.

\begin{figure}
    \includegraphics[width=0.5\textwidth]{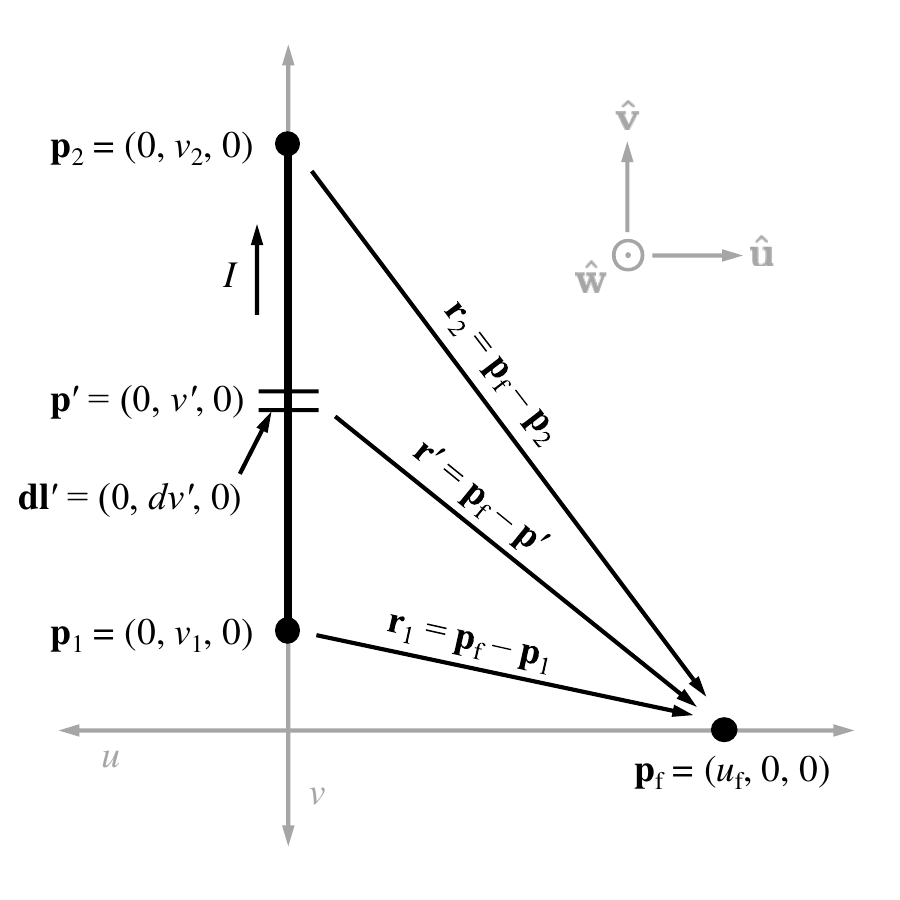}
    \caption{Setup for the computation of the Biot-Savart integral over a
             straight segment of wire.}
    \label{fig:seg_B_diagram}
\end{figure}

To compute the integral, the wireframe and test point will be placed in the
coordinate system $(u,v,w)$ shown in Fig.~\ref{fig:seg_B_diagram}. The 
evaluation point lies on the $u$ axis at the point $(u_f,0,0)$ and the segment
is positioned on the $v$ axis between the points $(0,v_1,0)$ and $(0,v_2,0)$.
The differential interval $\mathbf{dl}'$ along the segment is $(0,dv',0)$. The
vector $\mathbf{r}'$ is then $(u_f,0,0) - (0,v',0) = (u_f,-v',0)$ and


\begin{align}
    \begin{split}
        \mathbf{dl}' \times \mathbf{r}' &= (0, 0, -u_f dv') \\
                                        &= -u_f dv' \mathbf{\hat{w}}
    \end{split}
\end{align}

\noindent Here, $\mathbf{\hat{w}}$ is a unit vector perpendicular to the plane
containing the segment and the evaluation point (defined more precisely below).

The integral may then be written as

\begin{align}
    \label{eqn:bs_uvw}
    \mathbf{B} 
        &= - \frac{\mu_0 I}{4 \pi} \mathbf{\hat{w}}
            \int_{v_1}^{v_2} \frac{u_f}{\left(u_f^2 + v'^2\right)^{3/2}} dv'
\end{align}

\noindent and evaluates to 

\begin{align}
    \mathbf{B} 
        &= - \frac{\mu_0 I}{4 \pi} \left( \frac{1}{u_f} \right)
            \left( \frac{v_2}{\sqrt{u_f^2+v_2^2}} 
                   - \frac{v_1}{\sqrt{u_f^2+v_1^2}} \right)
            \mathbf{\hat{w}}
    \label{eqn:B_seg_uvw}
\end{align}


The expression in Eq.~\ref{eqn:B_seg_uvw} may be rewritten in coordinate-free
form by noting the following. First, the unit vectors $\mathbf{\hat{u}}$,
$\mathbf{\hat{v}}$, and $\mathbf{\hat{w}}$ can be expressed in terms of 
$\mathbf{r}_\text{1}$ and $\mathbf{r}_\text{2}$ as

\begin{align}
    \label{eqn:what}
    \mathbf{\hat{w}} &= \frac{\mathbf{r}_\text{2} \times \mathbf{r}_\text{1}}
        {\left|\mathbf{r}_\text{2} \times \mathbf{r}_\text{1}\right|} \\
    \label{eqn:vhat}
    \mathbf{\hat{v}} &= \frac{\mathbf{r}_\text{1} - \mathbf{r}_\text{2}}
        {\left|\mathbf{r}_\text{1} - \mathbf{r}_\text{2}\right|} \\
    \label{eqn:uhat}
    \mathbf{\hat{u}} &= \mathbf{\hat{v}} \times \mathbf{\hat{w}} 
        = \frac{
                \left( \left|\mathbf{r}_\text{2}\right|^2
                    - \mathbf{r}_\text{1} \cdot \mathbf{r}_\text{2} 
                    \right)
                \mathbf{r}_\text{1} + 
                \left( \left|\mathbf{r}_\text{1}\right|^2
                    - \mathbf{r}_\text{1} \cdot \mathbf{r}_\text{2} 
                    \right)
                \mathbf{r}_\text{2}
               }
               {
                \left|\mathbf{r}_\text{1} - \mathbf{r}_\text{2}\right|
                \left|\mathbf{r}_\text{1} \times \mathbf{r}_\text{2}\right|
               }
\end{align}



Using \Cref{eqn:what,eqn:vhat,eqn:uhat}, the coordinates $v_1$, $v_2$, and
$u_f$ may be written as

\begin{align}
    \label{eqn:v1}
    v_1 &= -\mathbf{r}_\text{1} \cdot \mathbf{\hat{v}}
        = \frac{
                \mathbf{r}_\text{1} \cdot \mathbf{r}_\text{2} -
                \left|\mathbf{r}_\text{1}\right|^2 
               }
               {\left|\mathbf{r}_\text{1} - \mathbf{r}_\text{2}\right|} \\
    \label{eqn:v2}
    v_2 &= -\mathbf{r}_\text{2} \cdot \mathbf{\hat{v}}
        = \frac{
                \left|\mathbf{r}_\text{2}\right|^2  -
                \mathbf{r}_\text{1} \cdot \mathbf{r}_\text{2}
               }
               {\left|\mathbf{r}_\text{1} - \mathbf{r}_\text{2}\right|} \\
    \label{eqn:uo}
    u_f &= \mathbf{r}_\text{1} \cdot \mathbf{\hat{u}}
        = \frac{
                \left|\mathbf{r}_\text{1}\right|^2
                \left|\mathbf{r}_\text{2}\right|^2
                - \left(\mathbf{r}_\text{1} \cdot \mathbf{r}_\text{2}\right)^2
               }
               {
                \left|\mathbf{r}_\text{1} - \mathbf{r}_\text{2}\right|
                \left|\mathbf{r}_\text{1} \times \mathbf{r}_\text{2}\right|
               }
        = \frac{
                \left|\mathbf{r}_\text{1} \times \mathbf{r}_\text{2}\right|
               }
               {
                \left|\mathbf{r}_\text{1} - \mathbf{r}_\text{2}\right|
               }
\end{align}

Finally, substituting \Cref{eqn:what,eqn:v1,eqn:v2,eqn:uo} into 
\Cref{eqn:B_seg_uvw}, also noting that 
$\sqrt{u_f^2+v_1^2}=\left|\mathbf{r}_1\right|$ and
$\sqrt{u_f^2+v_2^2}=\left|\mathbf{r}_2\right|$, yields, after some algebra,
the expression in Eq.~\ref{eqn:segment_Bfield}:

\begin{align}
\label{eqn:segment_Bfield_recap}
    \mathbf{B}
        &= \frac{\mu_0 I}{4\pi}
           \left(\frac{|\mathbf{r}_1| + |\mathbf{r}_2|}
                    {|\mathbf{r}_1||\mathbf{r}_2|
                        (|\mathbf{r}_1||\mathbf{r}_2| 
                             + \mathbf{r}_1 \cdot \mathbf{r}_2)} \right)
           \mathbf{r}_1 \times \mathbf{r}_2
\end{align}

The expression is mathematically valid for any point that does not lie on the 
segment between $\mathbf{p}_\text{1}$ and $\mathbf{p}_\text{2}$, in which case 
the term $|\mathbf{r}_\text{1}||\mathbf{r}_\text{2}| + \mathbf{r}_\text{1}
\cdot \mathbf{r}_\text{2}$ in the denominator is zero. This situation should 
not arise in practical wireframe applications, however, as the magnetic field
of interest is located on the plasma boundary, which is assumed not to 
intersect any wireframe segment.

In general, using a filamentary approximation for a wire with finite thickness 
introduces an error in the magnetic field calculation proportional to 
$(t/d)^2$, where $t$ is the thickness of the wire and $d$ is the distance from 
the wire to the
field evaluation point \cite{mcgreivy2021a}. In principle, this error could
be reduced with a more sophisticated model for the wireframe in which 
each segment has a finite thickness and the junction geometry at the nodes
is implemented carefully such that the continuity constraint equations
(Eq.~\ref{eqn:constr_continuity}) still prevent charge accumulation.
However, related to the discussion Sec.~\ref{sec:gsco_method}, even with
finite segment build, the wireframe current distribution is unlikely to 
resemble a final coil design. Rather, the wireframe current distribution
would need to be refined such that the current flows take smoother paths.
It may more worthwhile to account for finite thickness in this subsequent
refining stage rather than during the optimization of the wireframe currents.


\section{Derivation of the equations for RCLS}
\label{apx:rcls_derivation}


The RCLS problem in Eq.~\ref{eqn:rcls_problem} constitutes a linear 
least-squares minimization subject to linear equality constraints and a
regularization term. The first step in solving the problem is to 
represent the solution space for
$\mathbf{x}$ in a basis in which its components are either fully
determined by the constraints (i.e.~in the image of the constraint matrix)
or completely unconstrained (i.e.~in the null space of the constraint matrix). 
As described, for example, in Ref.~\cite{bjoerck2024a} (Section 3.4) or 
\cite{golub2005a}, this can be done
by computing the ``full'' QR factorization of the transpose of the constraint 
matrix $\mathbf{C}^T$:

\begin{equation}
    \label{eqn:qr_fact}
    \mathbf{C}^T = \begin{bmatrix} \mathbf{Q} & \mathbf{\tilde{Q}} \end{bmatrix}
                   \begin{bmatrix} \mathbf{R} \\ 0 \end{bmatrix}
\end{equation}

\noindent Here, $\mathbf{C}\in\mathbb{R}^{p \times n}$ is the constraint matrix as defined in Eq.~\ref{eqn:constr_all}, $n$ is the number of elements in 
$\mathbf{x}$ (the number of wireframe segments), and $p$ is the number of 
linearly independent constraint equations. The $p$ columns of submatrix 
$\mathbf{Q}\in\mathbb{R}^{n \times p}$ and the $n-p$ columns of 
$\mathbf{\tilde{Q}}\in\mathbb{R}^{n \times (n-p)}$ are orthonormal unit vectors,
and $\mathbf{R}\in\mathbb{R}^{(n-p) \times (n-p)}$ is an upper triangular 
matrix.

Now define vectors $\mathbf{u}\in\mathbb{R}^p$ and 
$\mathbf{v}\in\mathbb{R}^{n-p}$ in the bases specified by the columns of
$\mathbf{Q}$ and $\mathbf{\tilde{Q}}$:

\begin{equation}
    \label{eqn:cob}
    \begin{bmatrix} \mathbf{Q}^T \\ \mathbf{\tilde{Q}}^T \end{bmatrix}
        \mathbf{x}
            = \begin{bmatrix} \mathbf{u} \\ \mathbf{v} \end{bmatrix}
\end{equation}

\noindent Using the orthogonality property of the matrix on 
the left-hand side of Eq.~\ref{eqn:cob}, $\mathbf{x}$ may be written in terms
of $\mathbf{u}$ and $\mathbf{v}$ as

\begin{equation}
    \label{eqn:x_u_v}
    \mathbf{x} = \begin{bmatrix}\mathbf{Q}&\mathbf{\tilde{Q}}\end{bmatrix}
                 \begin{bmatrix}\mathbf{u}\\\mathbf{v}\end{bmatrix}
               = \mathbf{Q}\mathbf{u} + \mathbf{\tilde{Q}}\mathbf{v}
\end{equation}

Using Eq.~\ref{eqn:x_u_v}, Eq.~\ref{eqn:rcls_problem_constr} can be expressed 
as

\begin{equation}
    \label{eqn:constr_qr}
    \mathbf{R}^T \mathbf{Q}^T \mathbf{x} = \mathbf{d},
\end{equation}

\noindent or, after incorporating Eq.~\ref{eqn:cob},

\begin{equation}
    \label{eqn:constr_u_basis}
    \mathbf{R}^T \mathbf{u} = \mathbf{d}
\end{equation}

\noindent Thus, the constraints fix the vector $\mathbf{u}$ whereas $\mathbf{v}$
represents the part of the solution that is unrestricted by the constraints. 
Eq.~\ref{eqn:constr_u_basis} can be solved efficiently for 
$\mathbf{u}$ using forward substitution, as $\mathbf{R}^T$ is lower triangular.

With $\mathbf{u}$ now known, it remains to determine $\mathbf{v}$, the 
unconstrained part of the solution. Using Eq.~\ref{eqn:x_u_v}, the total
objective function $f$ may be written as

\begin{equation}
    \label{eqn:f_u_v}
    f = \frac{1}{2} \left( \mathbf{A}\mathbf{Q}\mathbf{u} + 
                           \mathbf{A}\mathbf{\tilde{Q}}\mathbf{v}
                           - \mathbf{b} \right)^2 + 
        \frac{1}{2} \left( \mathbf{W}\mathbf{Q}\mathbf{u} + 
                           \mathbf{W}\mathbf{\tilde{Q}}\mathbf{v} \right)^2
\end{equation}

The least-squares solution $\mathbf{v}^*$ that minimizes $f$ is that which
causes the gradient of $f$ to vanish:

\begin{equation}
    \label{eqn:grad_f_v}
    \nabla_\mathbf{v} f =
        \mathbf{\tilde{Q}}^T \mathbf{A}^T \left( 
            \mathbf{A} \mathbf{Q} \mathbf{u} +
            \mathbf{A} \mathbf{\tilde{Q}} \mathbf{v} - \mathbf{b} \right) +
        \mathbf{\tilde{Q}}^T \mathbf{W}^T \left( 
             \mathbf{W} \mathbf{Q} \mathbf{u} +
             \mathbf{W} \mathbf{\tilde{Q}} \mathbf{v} \right)
        = 0
\end{equation}

Collecting like terms and rearranging leads to

\begin{equation}
    \label{eqn:Av=b}
    \mathbf{\tilde{Q}}^T \left(
        \mathbf{A}^T \mathbf{A} \mathbf{\tilde{Q}} +
        \mathbf{W}^T \mathbf{W} \mathbf{\tilde{Q}} \right) \mathbf{v}
    = 
    \mathbf{\tilde{Q}}^T \left(
        \mathbf{A}^T \mathbf{b} - \mathbf{A}^T \mathbf{A} \mathbf{Q} -
        \mathbf{W}^T \mathbf{W} \mathbf{Q} \right) \mathbf{u} 
\end{equation}

Eq.~\ref{eqn:Av=b} constitutes a system of linear equations that can then be 
solved directly for $\mathbf{v}$. Along with
the solution for $\mathbf{u}$ determined in Eq.~\ref{eqn:constr_u_basis},
the solution $\mathbf{x}^*$ to the constrained least squares problem can
then be computed via Eq.~\ref{eqn:x_u_v}.

For high-resolution wireframes, in which the matrices $\mathbf{A}$ and 
$\mathbf{W}$ are large, the problem may be difficult to solve directly
due to memory constraints and poor matrix conditioning. In this case,
iterative algorithms with preconditioners may be employed. Similar
high-dimensional constrained linear least squares problems have been solved
with such techniques, for example, in Ref.~\cite{kaptanoglu2023a}.


\section{Dependence of GSCO solutions on wireframe geometry}
\label{apx:wframe_geometry}

In each optimization shown in this paper, the free parameters were the currents
in each segment of the underlying wireframe. The positions, lengths, and 
connectivity of the wireframe segments were held fixed during the optimizations.
However, these geometric aspects naturally influence the ability of 
the optimizer to find an accurate solution.
To get an indication of the sensitivity of the solutions to the geometry of
the wireframe, a GSCO procedure was run on two wireframes with slight 
geometric differences. 

The first test wireframe was used for the optimizations performed in
Sec.~\ref{sec:gsco}. It was created by first generating a toroidal surface 
offset roughly 30~cm from the target plasma boundary. This surface was 
calculated by the \textsc{Bnorm} code. \textsc{Bnorm} 
was originally developed to prepare winding surfaces and calculate the target 
normal field on the plasma boundary for \textsc{Nescoil} computations 
\cite{merkel1987a}, and is currently available in the \textsc{Stellopt} code 
suite \cite{stellopt}. Its method for generating offset surfaces is 
described in Appendix B of Ref.~\cite{landreman2017a}.

The offset surface geometry is parametrized with a toroidal angle $\zeta$ and
a poloidal angle $\theta$. The toroidal angle $\zeta$ coincides with the 
standard azimuthal angle $\phi$ in cylindrical coordinates. The cylindrical 
radial ($r$) and vertical ($z$) coordinates of a point on the surface may be 
computed for a given $\theta$ and $\zeta$.

With this surface generated, the nodes of the wireframe were positioned on
the surface on a two-dimensional grid of evenly-spaced intervals of $\zeta$ 
and $\theta$. The resolution of the grid was $96 \times 100$, meaning 96
nodes per half-period in the toroidal dimension and 100 nodes in the poloidal 
dimension. 
Because of the equivalence of $\zeta$ with the cylindrical
azimuthal angle $\phi$ mentioned above, nodes with a given value of $\zeta$ 
all lie in a plane with the azimuthal angle $\phi=\zeta$. However, in the
poloidal dimension, the physical spacing between the nodes is, in general,
not regular.

Following the creation of the first wireframe, a second wireframe was generated
with slightly different geometry. The nodes were placed on the same toroidal
surface at the same azimuthal angles $\phi=\zeta$ and had the same grid 
resolution of $96 \times 100$; however, the poloidal
intervals between each set of nodes were adjusted such that, for any given 
$\zeta$, the Euclidean distances between nodes with adjacent $\theta$ values
were uniform. As a result, each segment in the second wireframe had a 
slightly different position and orientation than the corresponding segment
in the first wireframe with the same toroidal and poloidal index. 
Similarly, a given current in a particular segment of the second wireframe
would have a slightly different contribution to the magnetic field than the 
same current in the corresponding segment of the first wireframe; in other
words, the elements of the matrix $\mathbf{A}$ (Eq.~\ref{eqn:A_matrix_element})
would be different.

Both of these wireframes were initialized with six planar poloidal current
flows per half-period at identical toroidal angles. A GSCO procedure
was then run on each of the initialized wireframes. Both optimizations used
the same hyperparameters, including $\lambda_S=10^{-6}$ T$^2$m$^2$ and a 
loop current equal to the current in the initial poloidal flows.

\begin{figure*}
    \begin{center}
    \includegraphics[width=\textwidth]{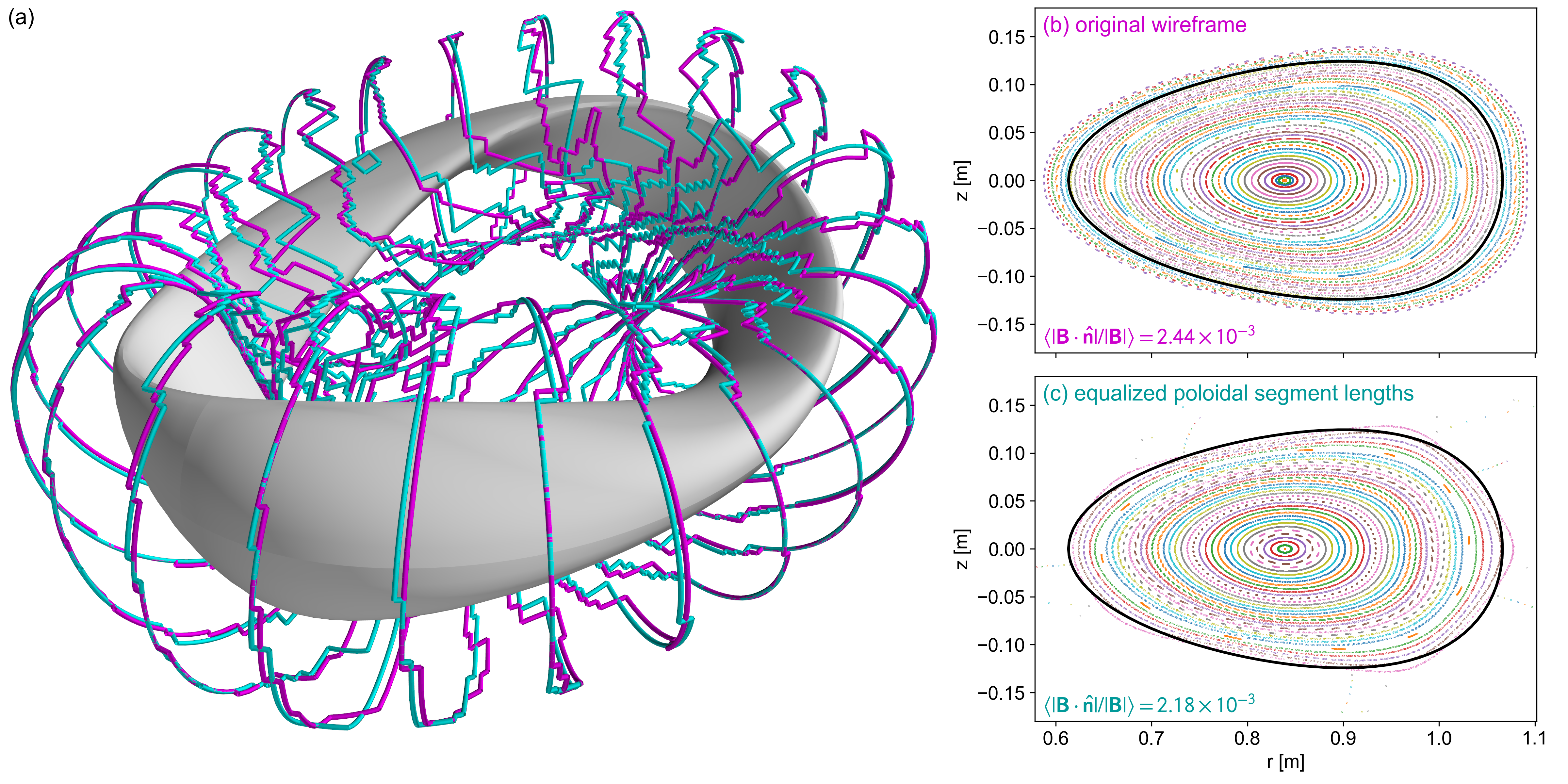}
    \caption{Comparison of GSCO solutions performed with the same 
             hyperparameters on wireframes with the same grid resolution but
             slightly different segment geometry.
             (a) Rendering of the active segments from the solution with the
                 original wireframe (magenta) and the wireframe with adjusted                    poloidal segment lengths (cyan), along with the target
                 plasma boundary;
             (b-c) Poincar\'e cross-sections of field lines traced for each
                 solution alongside the cross-section of the target plasma
                 boundary.}
    \label{fig:gsco_compare_wframes}
    \end{center}
\end{figure*}

A comparison of the solutions and their field accuracies is shown in 
Fig.~\ref{fig:gsco_compare_wframes}. In Fig.~\ref{fig:gsco_compare_wframes}a,
the active segments from both solutions are overlain. The solution on the
wireframe with the original geometry is shown in magenta, whereas the solution
on the wireframe with equalized poloidal segment lengths is shown in 
cyan. Both solutions consist mostly of modular coils, similar to the solutions
in Sec.~\ref{sec:gsco_modular} (in fact, the original solution is the 
solution from Sec.~\ref{sec:gsco_modular} with $\lambda_S=10^{-6}$ T$^2$m$^2$).
Overall, the modular coils follow qualitatively similar paths, especially
on the outboard side. However, some differences in geometry can be seen on
the inboard side. These differences are partly due to the different 
orientations of the segments in each wireframe; in some areas, it is 
clear that the segments from the modified wireframe are tilted at a substantial
angle relative to the segments in the original wireframe.

Overall, both solutions had very similar field accuracy metrics. The 
mean relative normal field $\mrbn$ was $2.44 \times 10^{-3}$ for the 
solution on the original wireframe, while the solution on the new wireframe
had a slightly lower value of $\mrbn = 2.18 \times 10^{-3}$.
However, the flux surfaces created by each solution have more notable
differences, as shown in the Poincar\'e cross-sections
in Fig.~\ref{fig:gsco_compare_wframes}b for the original wireframe
geometry and Fig.~\ref{fig:gsco_compare_wframes}c for the modified wireframe
geometry. Even though the solution on the modified wireframe had a slightly
better summary field accuracy metric $\mrbn$, it appears to be impacted
by a resonant field error that deforms the flux surface geometry near the 
target plasma boundary.

It is well understood that even a small error field can substantially deform or 
otherwise impact flux surface geometry in the vicinity of a surface with a
rotational transform value that is resonant with the error field 
\cite{rosenbluth1966a}. Thus, it is not unexpected that small changes in the
wireframe geometry might result in a field error that is small but large
enough to impact the flux surfaces. 

It is not ideal that the achievable flux surface quality of GSCO solutions can 
be so sensitive to small changes in the geometry of the underlying wireframe.
However, it is hypothesized that there is room for improvement in this
regard. For example, while the objective function $f_{GSCO}$ employed in these
optimizations includes a general metric of field accuracy, it does not
include specific, elevated penalties for resonant error fields. If such a
penalty were added, it might be possible to achieve a better solution with
the second wireframe. Alternatively, if the solution on the second wireframe
is refined with a space curve optimizer, this subsequent optimization could
use a resonant error objective in order to heal the resonant error field.
In the meantime, when developing wireframe solutions for a given plasma 
equilibrium, it is generally good practice to repeat optimizations with 
slight variations in wireframe geometry to see if better solutions are
achievable.

\end{document}